\documentclass[structabstract,USenglish]{aa}
\usepackage{graphicx}
\usepackage{txfonts}
\usepackage{longtable}
\usepackage[labelfont=bf]{caption}
\usepackage{lscape,supertabular}
\usepackage{natbib}
\usepackage{hyperref}
\usepackage{subfig}
\usepackage{enumitem}
\usepackage{inputenc}
\usepackage[range-units = brackets,tophrase={-},seperr,repeatunits=false]{siunitx}
\begin{document}

	\title{XMM-Newton and Chandra Cross Calibration\\Using HIFLUGCS Galaxy Clusters: \\Systematic Temperature Differences and Cosmological Impact}

   \author{G. Schellenberger\inst{\ref{inst1}}\thanks{gerrit@uni-bonn.de; member of the International Max Planck Research School (IMPRS) for Astronomy and Astrophysics at the Universities of Bonn and Cologne}
          ,
          T. H. Reiprich\inst{\ref{inst1}}
          ,
          L. Lovisari\inst{\ref{inst1}}
          ,
          J. Nevalainen\inst{\ref{inst2}, \ref{inst3}}
          \and
          L. David\inst{\ref{inst4}}
          }

   \institute{Argelander-Institut f\"ur Astronomie, Universit\"at Bonn, Auf dem
               H\"ugel 71, 53121 Bonn, Germany\label{inst1}
               \and
             Tartu Observatory, 61602 Toravere, Estonia\label{inst2}
             \and
             Department of Physics, Dynamicum, PO Box 48, 00014 University of Helsinki, Finland\label{inst3}
             \and
             Harvard-Smithsonian Center for Astrophysics, 60 Garden Street,
             Cambridge, USA\label{inst4}}

  \abstract
   {Robust X-ray temperature measurements of the
   intracluster medium of galaxy clusters require an accurate
   energy-dependent effective area calibration. Since the hot gas X-ray
   emission of galaxy clusters does not vary on relevant time scales they
   are excellent cross-calibration targets. Moreover, cosmological
   constraints from clusters rely on accurate gravitational mass
   estimates, which in X-rays strongly depend on cluster gas temperature
   measurements. Therefore, systematic calibration differences may result
   in biased, instrument-dependent cosmological constraints. This is of special interest in the light of the tension between the Planck results of the primary temperature anisotropies of the CMB and Sunyaev-Zel'dovich plus X-ray cluster counts analyses.}
   {We quantify in detail the systematics and uncertainties of the cross-calibration of the effective area between five X-ray instruments, EPIC-MOS1/MOS2/PN onboard XMM-Newton and ACIS-I/S onboard Chandra, and the influence on temperature measurements. Furthermore, we assess the impact of the cross calibration uncertainties on cosmology.}
  {Using the HIFLUGCS sample, consisting of the 64 X-ray brightest galaxy clusters, we constrain the ICM temperatures through spectral fitting in the same, mostly isothermal, regions and compare the different instruments. 
     We use the stacked residuals ratio method to evaluate the cross calibration uncertainties between the instruments as a function of energy. Our work is an extension to a previous one using X-ray clusters by the International Astronomical Consortium for High Energy Calibration (IACHEC) and is carried out in the context of IACHEC.}
  {Performing spectral fitting in the full energy band, $\SIrange{0.7}{7}{keV}$, as is typical for the analysis of cluster spectra, we find that best-fit temperatures determined with XMM-Newton/EPIC are significantly lower than Chandra/ACIS temperatures. This confirms with high precision the previous IACHEC results obtained with older calibrations. The difference increases with temperature and we quantify this dependence with a fitting formula. For instance, at a cluster temperature of $\SI{10}{keV}$, EPIC temperatures are on average 23\% lower than ACIS temperatures. We also find systematic differences between the three XMM-Newton EPIC instruments with the PN detector typically estimating the lowest temperatures. Testing the cross-calibration of the energy-dependence of the effective areas in the soft and hard energy bands, $\SIrange{0.7}{2}{keV}$ and $\SIrange{2}{7}{keV}$, respectively, we confirm the previously indicated relatively good agreement between all instruments in the hard and the systematic differences in the soft band. We provide scaling relations to convert between the different instruments based on the effective area, gas temperature and hydrostatic mass.
  We demonstrate that effects like multi temperature structure and different relative sensitivities of the instruments at certain energy bands cannot explain the observed differences. We conclude that using XMM-Newton/EPIC, instead of Chandra/ACIS to derive full energy band temperature profiles for cluster mass determination results in an 8\% shift towards lower $\Omega_{\rm M}$ values and $<1$\% change of $\sigma_8$ values in a cosmological analysis of a complete sample of galaxy clusters. Such a shift alone is insufficient to significantly alleviate the tension between Planck cosmic microwave background primary anisotropies and Sunyaev-Zel'dovich plus XMM-Newton cosmological constraints.}
   {}
\titlerunning{XMM-Newton and Chandra Cross Calibration}
\authorrunning{Schellenberger et al.}
\date{Accepted by A\&A}
   \maketitle

\section{Introduction}\label{sec:intro}
Galaxy clusters are excellent tools to study cosmology, in particular the phenomena of dark matter and dark energy, because they are the most massive gravitationally relaxed systems in the Universe. Especially the cluster mass function, i.e. the number density of clusters with a certain mass, is a sensitive probe of cosmological parameters. By determining the temperature using X-ray emission from the hot intracluster medium (ICM) one traces the most massive visible component of clusters and can derive the total gravitating mass and also the mass of the emitting medium. 

Robust cosmological constraints require accurate estimates of the cluster masses without any systematic bias. There are at least two important sources of possible biases in the hydrostatic method:
a) the previously reported results of the International Astronomical Consortium for High Energy Calibration IACHEC\footnote{\url{http://web.mit.edu/iachec/}} on the cross calibration uncertainties of the effective area between XMM-Newton/EPIC and Chandra/ACIS (\citealp{nevalainen_2010}, N10), the two major current X-ray missions and 
b) the hydrostatic bias (e.g., \citealp{2007ApJ...668....1N}) whereby a fraction of the total pressure in the ICM is of non-thermal origin, e.g. due to bulk motions. In the latter case, the assumption of the gas pressure balancing the gravity yields an underestimation of the total mass. 

Both biases may be affecting the recent Planck results (\citealp{2013arXiv1303.5076P,2013arXiv1303.5080P}) whereby the cosmological constraints driven by the primary temperature anisotropies of the cosmic microwave background radiation (CMB) and Sunyaev-Zel'dovich (SZ) analyses are not in agreement with each other. For the SZ analysis a relation between the Compton $Y$ parameter and the  mass for galaxy clusters derived using XMM-Newton data was used and even allowing for a possible hydrostatic mass bias factor in the  
range $\numrange{0.7}{1.0}$ no full  agreement  between the two probes is achieved.
As mentioned before, reasons for the discrepancy can be the break down of the hydrostatic assumption, the underestimation of calibration uncertainties in the X-ray, but also in the microwave regime, or an incomplete cosmological model (e.g. the lack of massive neutrinos).

As pointed out in \cite{2014arXiv1402.2670V} the XMM-Newton based cluster masses in the Planck sample are significantly lower compared to values obtained from a weak lensing analysis. This difference is mass dependent and might be explained by (i) a temperature dependent calibration uncertainty and/or (ii) a failing hydrostatic assumption. We will address the question whether a Chandra derived scaling relation could solve this tension, i.e. whether systematic cross-calibration uncertainties between Chandra and XMM-Newton can explain the inconsistent Planck results. The results in \cite{2014arXiv1402.3267I}, where Chandra X-ray masses are in agreement with cluster masses of a weak lensing analysis indicate that the hydrostatic assumption does not cause a major bias with respect to weak lensing masses. Note that while we will isolate here clearly the systematic uncertainty resulting directly and only from X-ray calibration uncertainties, a comparison of mass estimates and/or cosmological constraints from different sources is much more complicated (\citealp{2014MNRAS.438...78R}).

We show in this work how reliable the current (December 2012) calibrations are by using nearby galaxy clusters as reference objects and comparing the measured XMM-Newton temperatures with the results from Chandra. Galaxy clusters are Megaparsec-scale objects and their X-ray emission from the hot ICM does not vary on human timescales. 

The XMM-Newton/Chandra effective area cross calibration uncertainties as reported in N10 yielded that Chandra/ACIS measures $\sim 10-15\%$ higher temperatures in the $\SIrange{0.5}{7}{keV}$ energy band compared to XMM-Newton/EPIC.
Since the N10 sample was relatively small (11 galaxy clusters), it is important to evaluate the XMM-Newton/Chandra effective 
area cross-calibration uncertainties with a large cluster sample to gain more statistical precision for the comparison of XMM-Newton and Chandra temperatures.
Furthermore, the cosmological implications of cross-calibration uncertainties were not studied by N10.
In more recent works (e.g., \citealp{2013ApJ...767..116M,2014MNRAS.443.2342M,2014arXiv1405.7876D}) the authors still find significant differences in temperatures between Chandra and XMM-Newton by comparing the data of typically of the order of 20 galaxy clusters.

Here we use the HIFLUGCS cluster sample (\citealp{reiprich_hiflugcs}), which provides the 64 galaxy clusters with the highest X-ray flux. High quality Chandra/ACIS and XMM-Newton/EPIC data are available for all of them except one (\citealp{hudson_what_2009,zhang_hiflugcs:_2010}).
This work is an extension to N10 in the sense that it updates the calibration information (as in Dec 2012) with  $\sim$5 times more objects and that the cosmological impact is quantified.
 The current data can be used to evaluate both the energy dependence and the normalization of the cross-calibration uncertainties. We are interested in the cross-calibration effect on the cluster mass function, which depends on the temperature and only on the gradient of the gas density. Therefore, the normalization of the effective areas is not relevant for the current work and will be addressed in detail in an upcoming paper.

In this paper we first describe the properties of the galaxy clusters we use in Section \ref{sec:properties}. We give an overview of our data reduction for the two satellites in Section \ref{sec:data} and describe the background subtraction in Section \ref{sec:background} in more detail. 
Section \ref{sec:data_analysis} deals with the analysis method and finally we present the results in Section \ref{sec:results} and discuss the various effects that might have an influence on the results in Section \ref{sec:disc}. In this Section we also describe the cosmological impact on the normalized matter density parameter, $\Omega_{\rm m}$ and the amplitude of the linear matter power spectrum on $\num{8}h^{-1}\si{Mpc}$ scale, $\sigma_8$.

Throughout this paper we use a flat $\Lambda$CDM cosmology with the following parameters:\\$\Omega_{\rm m} = \num{0.3}$, $\Omega_\Lambda = \num{0.7}$, $H_0 =  h \cdot \SI{100}{km~s^{-1}~Mpc^{-1}}$ with $h = \num{0.71}$.

\section{Cluster properties}\label{sec:properties}
The whole HIFLUGCS cluster sample consists of 64 clusters with a redshift up to $z=\num{0.215}$ for RXCJ1504. The average redshift of this sample is $z = \num{0.053}$ with a dispersion of $\num{0.039}$. All clusters are listed in Table \ref{tab:hiflugcs}. Chandra ACIS data are available for all 64 clusters, XMM-Newton EPIC data for all except A2244, which will be observed in AO13.

In \cite{hudson_what_2009} a galaxy cluster is defined as a cool core cluster if the central cooling time is less than $\SI{7.7}{Gyr}$. The authors also show that NCC clusters of the HIFLUGCS sample exhibit a temperature drop towards the center of less than 20\%, while CC clusters show a decrease of the temperature by a factor of 2 to 5. More details on this phenomenon including the categorization of the HIFLUGCS clusters in CC and NCC clusters can be found in \cite{hudson_what_2009}. 
To minimize possible biases introduced by multi temperature ICM (Section \ref{sec:multi_icm}) we exclude the cool core regions of clusters as provided by \cite{hudson_what_2009}[Tab. 2].

The greatest limitation for the choice of the extraction region size is imposed by ACIS-S. In order not to loose a fraction of the cluster annulus due to relatively limited ACIS-S field of view (FOV), we set the outer extraction radius to $\SI{3.5}{\arcmin}$ around the emission peak as defined in \cite{hudson_what_2009}.

Finally we always added $\SI{15}{\arcsec}$ on all cool core radii and point source radii (as determined from Chandra data using a wavelet algorithm) to minimize the scattering of emission into the annulus because of the XMM-Newton PSF. We marked bad columns and chip gaps in the observations of the three EPIC instruments and excluded them from all spectral analyses.

Furthermore, this assures that regions are fully covered by Chandra's FOV for almost all clusters. 
All this leads to the following procedure:
\begin{itemize}
\item For cool core clusters the cool core region (plus $\SI{15}{\arcsec}$) was excluded and the temperature was measured within an annulus between the cool core and the $\SI{3.5}{\arcmin}$;
\item non-cool core clusters are assumed to have no big temperature variations, so the region used here is the full $\SI{3.5}{\arcmin}$-circle. We assume that the azimuthal temperature variations can be neglected for our purpose;
\item for five clusters the $\SI{3.5}{\arcmin}$ enclose regions outside the \mbox{ACIS-S} chips. Since this could bias our results we changed for these clusters the outer border (for both, XMM-Newton and Chandra) from the usual $\SI{3.5}{\arcmin}$ to a smaller radius which lies completely on the chips. The new outer borders are: 
\begin{itemize}
\item[\textbullet] $\SI{2}{\arcmin}$ for Abell 754
\item[\textbullet] $\SI{2}{\arcmin}$ for Abell 1367
\item[\textbullet] $\SI{2.1}{\arcmin}$ for Abell 2256
\item[\textbullet] $\SI{2.4}{\arcmin}$ for Hydra A
\item[\textbullet] $\SI{2.3}{\arcmin}$ for NGC 1550
\end{itemize}
\item for seven clusters\footnote{A2142, A2256, A3526, HydraA, NGC1399, NGC4636, NGC5044}, the cool core is larger than $\SI{3.5}{\arcmin}$. These clusters were excluded from our analysis. To see whether these clusters would bias our result, we analyzed them within the full $\SI{3.5}{\arcmin}$-circle as well (see Fig. \ref{fig:excluded}). The check revealed that these clusters do not show a special behavior in any direction as compared to the other clusters. Apart from the previously mentioned figure and the stacked residuals ratio analysis (Section \ref{ch:srr}), these seven clusters are always excluded.
\end{itemize}
Within these regions the source to background count rate ratio in the sample is between 9 and 135 in the $\SIrange{0.7}{2.0}{keV}$ band.

\section{Data processing}\label{sec:data}
\subsection{Chandra}\label{sec:data_chandra}
Chandra data reduction was performed using the \textit{CIAO} software (CIAO 4.5, CALDB 4.5.5.1) and the \textit{HEASOFT} tools (6.12) including the \textit{Xspec} fitting package (12.7.1d and AtomDB 2.0.2)\footnote{\url{http://www.atomdb.org}}. We first created the level 2 event files using the contributed script \verb+chandra_repro+, e.g.\ to correct for afterglows from cosmic rays. In the next step the chips comprising the selected regions (I-Chips are combined for I-Observations) were cleaned from solar flares by creating a lightcurve with the suggested values from Markevitch's Cookbook\footnote{\url{http://cxc.harvard.edu/contrib/maxim/acisbg/COOKBOOK}} and the \verb+lc_clean+ algorithm. Point sources detected by the \verb+wavdetect+ algorithm are excluded, but every observation was visually inspected for false and not significant detections and, if necessary, the list of point sources was edited manually. The \verb+acis_bkgrnd_lookup+ script helped us to find the blank-sky background file from the CALDB matching to the observation. 
Finally background files have been reprojected to match the orientation of the cluster observation, before continuing. For VFAINT observations only the events with status bits $= 0$ are used because all background files are taken from VFAINT observations. Unfortunately the quiescent particle background varies with time, so we have to compensate for this behavior by rescaling the background count rate with a normalization factor. This factor is the ratio of the count rates in the $\SIrange{9.5}{12}{keV}$ energy range of the blank sky files and the observations, because the effective area of Chandra is almost zero in this energy interval and almost all events are related to the particle background. This factor is then  multiplied to the BACKSCALE value of the background spectral file. To create the spectra and response files we use the \verb+specextract+ task and create the weighted RMFs and ARFs. The spectra are grouped to have at least 30 counts per bin. 

\subsection{XMM-Newton}\label{sec:data_xmm}
The XMM-Newton data reduction is in some parts different from the Chandra treatment. The software we used was the SAS package version 12.0.1 with CCF calibration files from 14.12.2012. With \verb+emchain+ and \verb+epchain+ we created the initial event lists excluding flagged events (\verb+FLAG==0+) and setting \verb+PATTERN<=12+ for MOS and \verb+PATTERN==0+ for PN. The decision, that only single events are selected for EPIC-PN is due to the fact that there are still significant gain problems for the double events (see XMM-Newton release note XMM-CCF-REL-309). Blank-sky background files are also available for XMM-Newton and can be downloaded from the website\footnote{\url{http://xmm2.esac.esa.int/external/xmm_sw_cal/background/blank_sky.shtml}}. One has to make sure to use the correct file for each observation in terms of used filter, pointing and recording mode. Solar flares are a tremendous issue for XMM, so we cleaned the lightcurve in two steps to remove them. To obtain the good time intervals we fitted a Poisson distribution function to a $\SI{100}{s}$ binned histogram of the high energy lightcurve, i.e. $\SIrange{10}{12}{keV}$ for MOS and $\SIrange{12}{14}{keV}$ for PN. Events belonging to a count rate higher than $2 \sqrt{\mu}$ above the mean count-rate, $\mu$, were rejected. In a second step the lightcurve was filtered in the full energy band from $\SIrange{0.3}{10}{keV}$ by 
the same method as described before. The same thresholds were applied to the background files.
We visually inspected all final full energy band lightcurves and, if necessary, removed strong flares that were not detected by the previously mentioned method.
Since Chandra has a much better spatial resolution, we used the detected point sources from the Chandra analysis and removed here the same regions. 
We excluded chip gaps and bad columns of the EPIC instruments from all observations. This exclusion criterion changes the best-fit temperature on the order of 0.5\%. 
The normalization of the background files due to the changing particle background level is done in the same way as for Chandra except the high energy interval, $\SIrange{9.5}{12}{keV}$ for Chandra, $\SIrange{10}{12}{keV}$ for EPIC-MOS and $\SIrange{12}{14}{keV}$ for EPIC-PN, is different because of Chandra's lower effective area in the high energy. For the EPIC spectra we use the same spectral grouping parameters as for Chandra of 30 counts per bin.

\section{Background treatment}\label{sec:background}
X-ray observations are always contaminated with events not related to the source, which we call background in the following. It usually consists of: 
\begin{itemize}
\item The particle background: Energetic particles produce charges on the detector while penetrating it, or induce fluorescent emission  lines in the surrounding material; 
\item the so-called soft protons: These particle events should be removed by the flare reduction process to a significant amount;
\item the cosmic X-ray background (CXB): It can be subdivided into the local hot bubble emission, the thermal emission of the galactic halo and the contribution of unresolved point sources (most likely AGNs, see e.g.\ \citealp{2006ApJ...645...95H});
\item solar wind charge exchange emission (SWCX): Highly ionized particles interacting with neutral atoms. For more details see \cite{2004ApJ...607..596W,2004ApJ...610.1182S}.
\end{itemize}
\begin{figure}[t]
  \resizebox{\hsize}{!}{\includegraphics[width=0.75\textwidth]{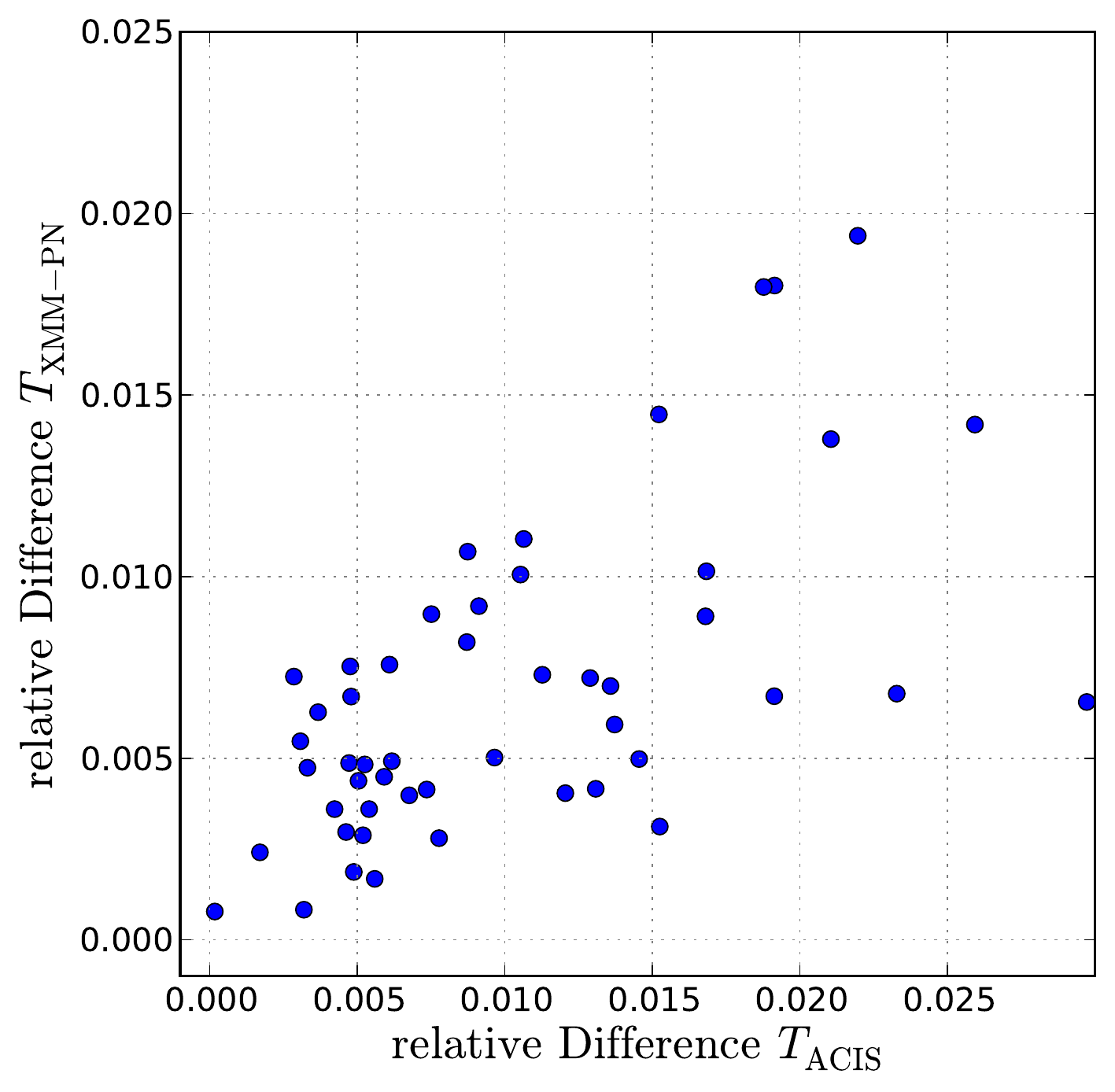}}
  \caption{Relative temperature difference resulting from raising the background level by 10 \% for energies below $\SI{2}{keV}$.}
  \label{fig:background}
\end{figure}
Generally it is of high importance to properly account for all background components as they affect the cluster temperature resulting from the spectral fit. The HIFLUGCS clusters are nearby objects, so they cover almost the whole field of view of the two instruments and the background cannot be estimated from the observation itself. We decided to use the blank-sky observations for both, Chandra and XMM-Newton, because our selection criteria ensures that the background is much lower than the source count rate (see Table \ref{tab:temp}). These archival observations of regions without astronomical sources are not always the best description for every observation, but when extracting spectra out to only $\SI{3.5}{\arcmin}$ for these very bright objects the error is negligible, because the background level is at least one order of magnitude lower than the cluster emission. To verify this assumption we increased the background normalization by $10\%$ up to an energy of $\SI{2}{keV}$. Beyond that energy the spectrum remains unchanged. This should simulate a different foreground/CXB emission level but leaves the particle background unchanged, which is dominant beyond $\SI{2}{keV}$.
For ACIS 91\% of the clusters exhibit a difference in the best-fit temperature of less than 2\%, while EPIC-PN temperature changes are less than 2\% for all clusters.
RXCJ1504 has a temperature difference for ACIS of almost 3\% after changing the background. Still this is not significant, since this cluster is one of the hottest in the sample and has a temperature difference between ACIS and EPIC-PN of more than 35\% in the full energy band (see Table \ref{tab:temp}).

Apart from a systematic under- or overestimation of the background level, an energy dependence of the background spectrum mismatch can be introduced by the photo-electric absorption of the blank sky background: The observations from which the blank sky background is extracted were taken at different sky positions and undergo different absorptions. The mean hydrogen column density along the line of sight, $N_{\rm H}$, of all these regions will not be in agreement with most of the cluster observation. 
We checked for all the clusters the effect of changing the background spectra by absorbing it according to the cluster $N_{\rm H}$ minus the exposure weighted average Hydrogen column density of the background file. This method produces a background spectrum as it would have been observed through the cluster line-of-sight absorption. For all clusters the temperature difference is below 3\% and for almost 90\% of the clusters it is below 1\%. This is also expected since the source to background count rate ratio in the relevant energy band is very high.

In summary, our results are robust against systematic uncertainties in the background estimation.

\section{Analysis}\label{sec:data_analysis}
\subsection{Spectral Fitting Method}
\begin{figure}
  \centering
  \resizebox{0.95\hsize}{!}{\includegraphics{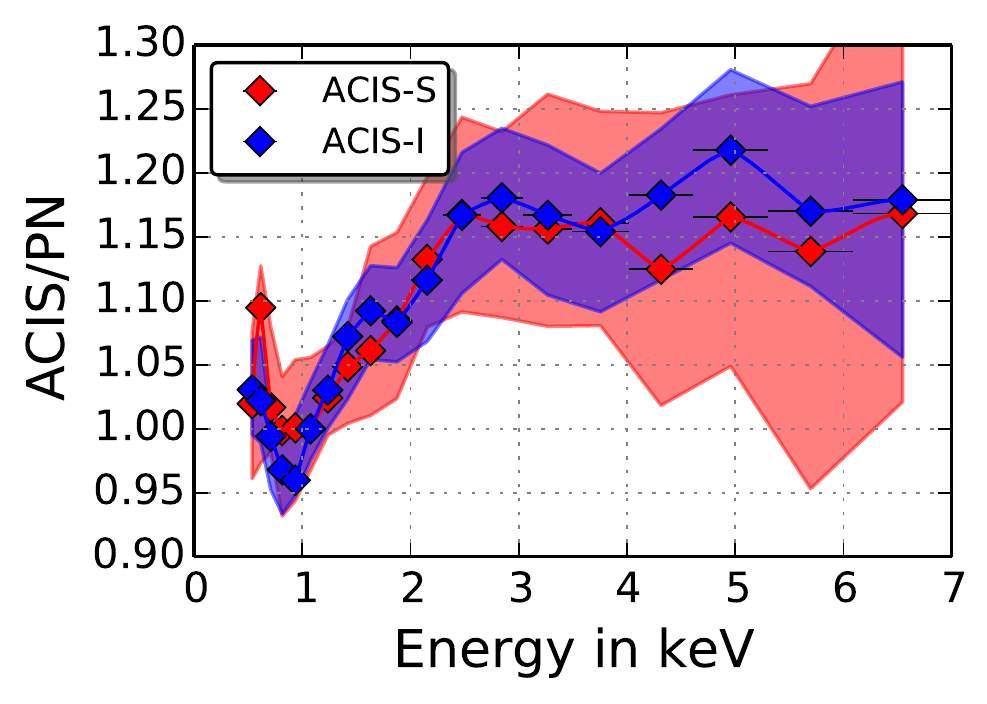}}
  \caption{Stacked residuals ratio, rescaled to unity at $\SI{1.1}{keV}$. The shaded region represents the 68\% confidence level from $\num{10000}$ bootstrap simulations. For a detailed description see text.}
  \label{fig:stacked}
\end{figure}
To obtain the emission model and temperature for each region we fit an \textit{apec}-model (\citealp{2001ApJ...556L..91S}) and a photoelectric absorption model (\textit{phabs}) using the cross sections from \cite{1992ApJ...400..699B}. 
The column density of neutral hydrogen $N_{\rm H}$ and the redshift $z$ are frozen to the values from \cite{zhang_hiflugcs:_2010}, where most of the hydrogen column densities are consistent with the values from the LAB HI survey (\citealp{2005A&A...440..775K}).
For two clusters, Abell 478 and Abell 2163, our spectral fits resulted in a very poor $\chi^2$, 
so we used $N_{\rm H}$ values from spectral fits with the hydrogen column density left free to vary. For Abell 478 we used $\SI{3e21}{cm^{-2}}$ and for Abell 2163 $\SI{2e21}{cm^{-2}}$. Both values are around 100\% higher than the LAB values and produced a $\chi_{\rm red}^2 < 1.6 $ for the spectral fit. The new values are also in rough agreement with the $N_{\rm H, tot}$ values, discussed in Section \ref{sec:nhtot}.

For Abell 478 and Abell 3571 we did not consider the redshifts from \cite{zhang_hiflugcs:_2010}, because all X-ray instruments are well in agreement with the new redshift values from the spectral fit (0.0848 instead of 0.0900 for Abell 478 and 0.0374 instead of 0.0397 for Abell 3571).

By default the abundance table presented in \cite{1989GeCoA..53..197A} (AnGr) was used for the absorption and emission model. Additionally all values were recomputed using the relative abundance of elements from \cite{2009ARA&A..47..481A} (Aspl). 

We performed the spectral fits in the full $\SIrange{0.7}{7}{keV}$, hard $\SIrange{2}{7}{keV}$ and soft  $\SIrange{0.7}{2}{keV}$ energy bands. This is different from the definition in N10, where the low energy threshold of the full and soft band was at $\SI{0.5}{keV}$. We excluded also the events below $\SI{0.7}{keV}$ to avoid the emission lines around $\SI{0.6}{keV}$ due to the Cosmic X-ray Background or SWCX.
\begin{figure*}
  \centering
  \resizebox{0.85\hsize}{!}{\includegraphics{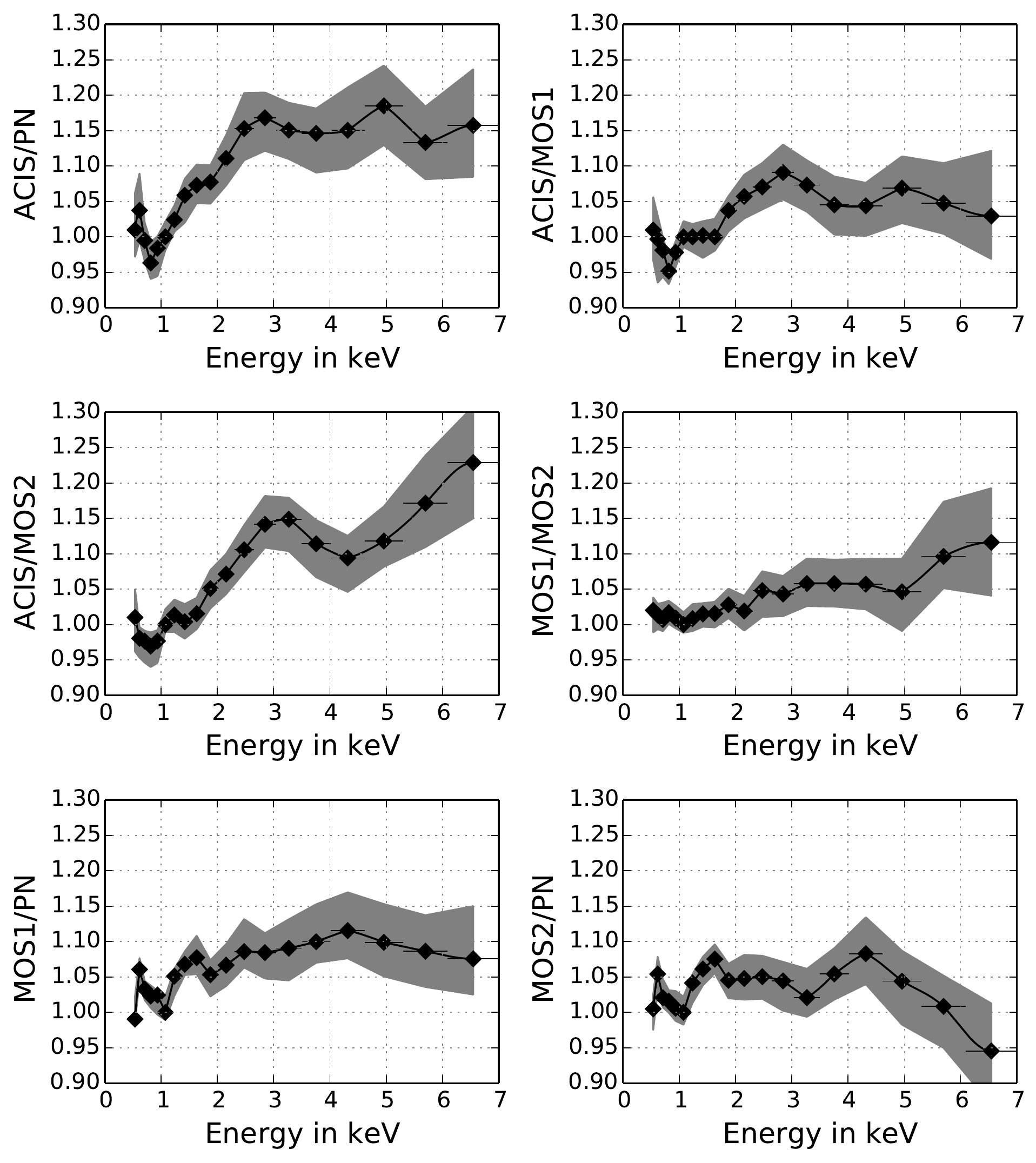}}
  \caption{Stacked residuals ratios are rescaled to unity at $\SI{1.1}{keV}$. The shaded region represents the 68\% confidence level from $\num{10000}$ bootstrap simulations. For a detailed description see text. Apart from choosing EPIC-PN as the reference instrument, we also show the ACIS/MOS1, ACIS/MOS2 and MOS1/MOS2 stacked residuals.}
  \label{fig:stacked2}
\end{figure*}
We also excluded all events above $\SI{7}{keV}$ because of the prominent EPIC-PN fluorescent lines (Ni at $\SI{7.5}{keV}$, Cu at $\SI{8}{keV}$ and Zn at $\SI{8.6}{keV}$) and the small Chandra effective area. It should be mentioned that the soft band spectral fits for high temperature clusters provides poor constraints on the parameters because the change in the slope of different temperature models is very small and there are almost no emission lines in this energy band.
For the full and hard band fits all $\chi_{\rm red}^2$ are below $\num{1.6}$, while for the soft band fits the maximum $\chi_{\rm red}^2$ is $\num{2.3}$ (although for more than $88 \%$ the $\chi_{\rm red}$ are below $\num{1.3}$).
The possibility and influence of a temperature-abundance degeneracy will be discussed in Section \ref{ch:tempabun}

\subsection{Stacked residuals ratio}\label{ch:srr}
The effective area cross-calibration uncertainties between a pair of instruments as a function of energy  can be obtained using the stacked residuals ratio method (see  \citealp{2013arXiv1301.2947K} and \citealp{2008RMxAC..32...62L}). Using the definition from \cite{2013arXiv1301.2947K} for the stacked residuals ratio,
\begin{equation}
R_{ij} = \frac{{\rm data}_{i}}{{\rm model}_j \otimes {\rm response}_i} \times \frac{{\rm model}_j \otimes {\rm response}_j}{{\rm data}_{j}}~,
\label{eq:srr}
\end{equation}
we can test instrument $i$ and use instrument $j$ as reference.
We briefly summarize the steps:
\begin{itemize}
\item The results on the relative cross-calibration uncertainties do not depend on the choice of the reference detector.
\item The data of the reference detector are fitted with an absorbed \verb+apec+-model in the $\SIrange{0.5}{7.0}{keV}$ energy range. We also experimented by using an absorbed \verb+mekal+-model to be more consistent with previous works (\citealp{2013arXiv1301.2947K}, N10) using the stacked residuals ratio method, but we did not find significant differences. This fitting procedure defines our reference model and is not changed any more in the following. Note that we included the $\SIrange{0.5}{0.7}{keV}$ energy band for the stacked residuals ratio test in order to stay close to previous analyses.
\item Each spectral data point of all the instruments (ACIS-S, ACIS-I, MOS1, MOS2 and PN) is divided by the reference model folded with the response (RMF and ARF) of the current instrument. This gives us the residuals.
\item In the next step we calculated the residuals ratio (Eq. \ref{eq:srr}) for each cluster and instrument pair by dividing the residuals of ACIS-S, ACIS-I,  MOS1 and MOS2 by the residuals of the reference instrument (PN).
The second term in Eq. \ref{eq:srr}  corrects for possible deviations between the data and model prediction of the reference instrument, i.e. after this correction the reference model agrees with the reference data. Thus, the details of the reference model (e.g. single temperature or multi temperature) are not important. Since the detectors use a different energy binning, we perform a linear interpolation to be able to calculate the residuals ratio at exactly the same energies for different instruments before dividing the residuals.
\item In the full $\SIrange{0.5}{7.0}{keV}$ energy range we define 19 energy bins (with equal separation in log-space), within which we calculate the median value of the residuals ratio of all the clusters to be analyzed. 
This yields the stack residuals ratio. We estimated the uncertainty by performing $\num{10000}$ bootstrap simulations and taking the 68\% confidence interval around the median of the $\num{10000}$ sample medians.
\item We normalized the stacked residuals ratios to unity at $\SI{1.1}{keV}$ because we are not studying the cross-calibration of the normalization of the effective area in this work since it does not affect the temperature measurement.
\end{itemize}
Applying the stacked residuals ratio method we end up with 53 clusters. Some observations\footnote{A262, A1367, A2029, A2052, A2589, A2634, A3571, A3581, A4059, NGC1550} had to be discarded because the source region is not completely within one chip for ACIS-S observations.

\section{Results}\label{sec:results}

\subsection{Stacked residuals ratio}\label{ch:stackedresiduals}
The analysis yielded that ACIS-S and ACIS-I stack residuals, using EPIC-PN as the reference, are consistent at all energies (see Fig. \ref{fig:stacked}).
This indicates that there are no energy-dependent effective area calibration biases between the two ACIS instruments, at the 5-10\% level of the statistical uncertainties. Thus, in the following we combine ACIS-S and ACIS-I to a single group called ACIS. The uncertainties coming from the bootstrap simulation are on average 60\% higher for ACIS-S/PN than for ACIS-I/PN, because only 13 observations are done with ACIS-S.

Furthermore, the flat ACIS and MOS1 v.s. PN stacked residuals imply that at energies above 3 keV there are no significant energy-dependent 
effective area cross-calibration uncertainties between these instruments.
However, the MOS2/PN stacked residuals deviate significantly from a flat ratio (see Fig. \ref{fig:stacked2}). Given that MOS2 is the only instrument indicating energy-dependent
features above 3 keV, it is likely that MOS2 has the larger calibration uncertainties, in the sense that with increasing energy in the 
3--7 keV band, the MOS2 effective area is increasingly overestimated. The bias reaches $\sim$10\% at 7 keV for MOS2/PN. 
Consequently, the MOS2 temperatures in the hard band are lower than values obtained with MOS1 or ACIS (see below).
A similar effect was suggested by N10, who did not study the effect in more detail due to lack of statistical precision.

At lower energies the situation is more problematic. In general, the EPIC-PN v.s. ACIS soft band differences are consistent with those reported in N10, but the better statistics and the systematic usage of stacked residual method in the present paper yield a more detailed view on the situation. ACIS and MOS1 v.s. PN stacked residuals exhibit a systematic decrease when moving from $\SI{3}{keV}$ to $\SI{1}{keV}$. The amplitudes are different:
$\sim$20\% for ACIS/PN and $\sim$10\% for MOS1/PN. The MOS2/PN and MOS1/MOS2 ratios have smaller ($\sim$5\%) and not significant changes when moving from $\SI{3}{keV}$ to $\SI{1}{keV}$. Especially in the $\SIrange{0.5}{2}{keV}$ band the MOS1/PN and MOS2/PN residuals show a very similar behavior.
If the shape of the effective area of PN was very accurately calibrated, the above results indicate that the ACIS (MOS1) effective area is 
overestimated by a factor of $\sim$20\% ($\sim$10\%) at $\SI{1}{keV}$.

The sudden increase of the stacked residuals ratio at energies $\SIrange{1}{0.5}{keV}$ by $\sim$5\% in all instruments compared to PN indicates problems with PN effective area calibration at these energies.
The simplest explanation of the data is that the  PN effective area is underestimated by 5\% at $\SI{0.5}{keV}$. We also detect this increase between $\sim\SI{1}{keV}$ and $\SI{0.5}{keV}$ in the ACIS/MOS1 and ACIS/MOS2 ratios but at lower level ($<5\%$).

While preparing this work, a paper by \cite{2014A&A...564A..75R} appeared on arXiv employing stacked residuals of on-axis point sources. While their results are mostly consistent with ours, there is an indication of a slightly different behavior in their default analysis method (\textit{stack+fit}) of the MOS2/PN case at high energies. They only see a drop in the stacked residuals for MOS2/PN, if they use the \textit{fit+stack} method, as we do here. In \cite{2014A&A...564A..75R} the authors mention ``negative spectral bins that can sometimes occur in the individual source spectra`` as the reason for the MOS2/PN drop in the \textit{fit+stack} method. We tested this by excluding all spectra from the stacked residuals analysis, which have at least one negative bin. Since no different behavior can be detected, we conclude that negative bins do not matter in our analysis.

\subsection{Temperatures}\label{ch:temps}
The results of the stacked residuals imply a multitude of temperature agreements and disagreements between different instruments in different energy bands (reported in Figs. \ref{fig:acisis} - \ref{fig:xmm_chan_high} and Table \ref{tab:temp}). In the following we will compare these temperatures and evaluate the significance of temperature differences for one cluster of the sample as well as for the whole sample.
\subsubsection{Temperature comparison of individual clusters}\label{ch:temps_first}
We evaluate the temperature differences and their significances $\xi$ by defining 
\begin{equation}
\xi = \frac{T_{I_X} - T_{I_Y}}{\sqrt{\Delta T_{I_X}^2 + \Delta T_{I_Y}^2}}~,
\end{equation}
where T$_{I_X} $ and T$_{I_Y} $ denote the temperatures measured with two instruments to be compared and  $\Delta T_{I_X}$ and $ \Delta T_{I_Y} $denote the statistical uncertainties of the temperature for the two instruments. Note that $\xi$ is calculated for each cluster and instrument combination individually. 
The $\xi$ distributions for the detector combinations are not symmetric, but we are able to calculate the median $\xi$ for each combination and the percentage of clusters with $\xi$ above 3 (see Table \ref{tab:significance}). For a non-systematic temperature difference (e.g. scatter) the median should not be significantly different from zero. 
\begin{table}[tbp]
\centering
\caption{Median of the significance of temperature differences, $\xi$, and probability of HIFLUGCS clusters to deviate more than 3 $\xi$ from zero for the three energy bands.}
\begin{tabular}{llcr}
Energy & Detector & Median & $P(\xi > 3)$ \\
Band & X--Y &  & \\
\hline
\multicolumn{1}{c}{Full} &ACIS--PN & 7.1 & 84\% \\
$\SIrange{0.7}{7.0}{keV}$&ACIS--MOS1 & 4.0 & 61\%  \\
&ACIS--MOS2 & 5.4 & 80\%  \\
&MOS1--MOS2 & 1.5 & 12\%  \\
&MOS1--PN & 2.3 & 34\% \\
&MOS2--PN & 1.3 & 14\% \\
\hline
\multicolumn{1}{c}{Soft} & ACIS--PN & 3.3 & 54\%  \\
$\SIrange{0.7}{2.0}{keV}$&ACIS--MOS1 & 1.5 & 16\% \\
&ACIS--MOS2 & 1.3 & 27\%  \\
&MOS1--MOS2 & 0.2 & 0\%  \\
&MOS1--PN & 1.6 & 20\% \\
&MOS2--PN & 1.4 & 20\% \\
\hline
\multicolumn{1}{c}{Hard} &ACIS--PN & 0.9 & 4\%  \\
$\SIrange{2.0}{7.0}{keV}$&ACIS--MOS1 & 0.9 & 2\% \\
&ACIS--MOS2 & 1.1 & 4\%  \\
&MOS1--MOS2 & 0.6 & 0\%  \\
&MOS1--PN & 0.5 & 2\% \\
&MOS2--PN & -0.1 & 2\% \\
\end{tabular}
\label{tab:significance}
\end{table}

\begin{figure}
  \resizebox{\hsize}{!}{\includegraphics{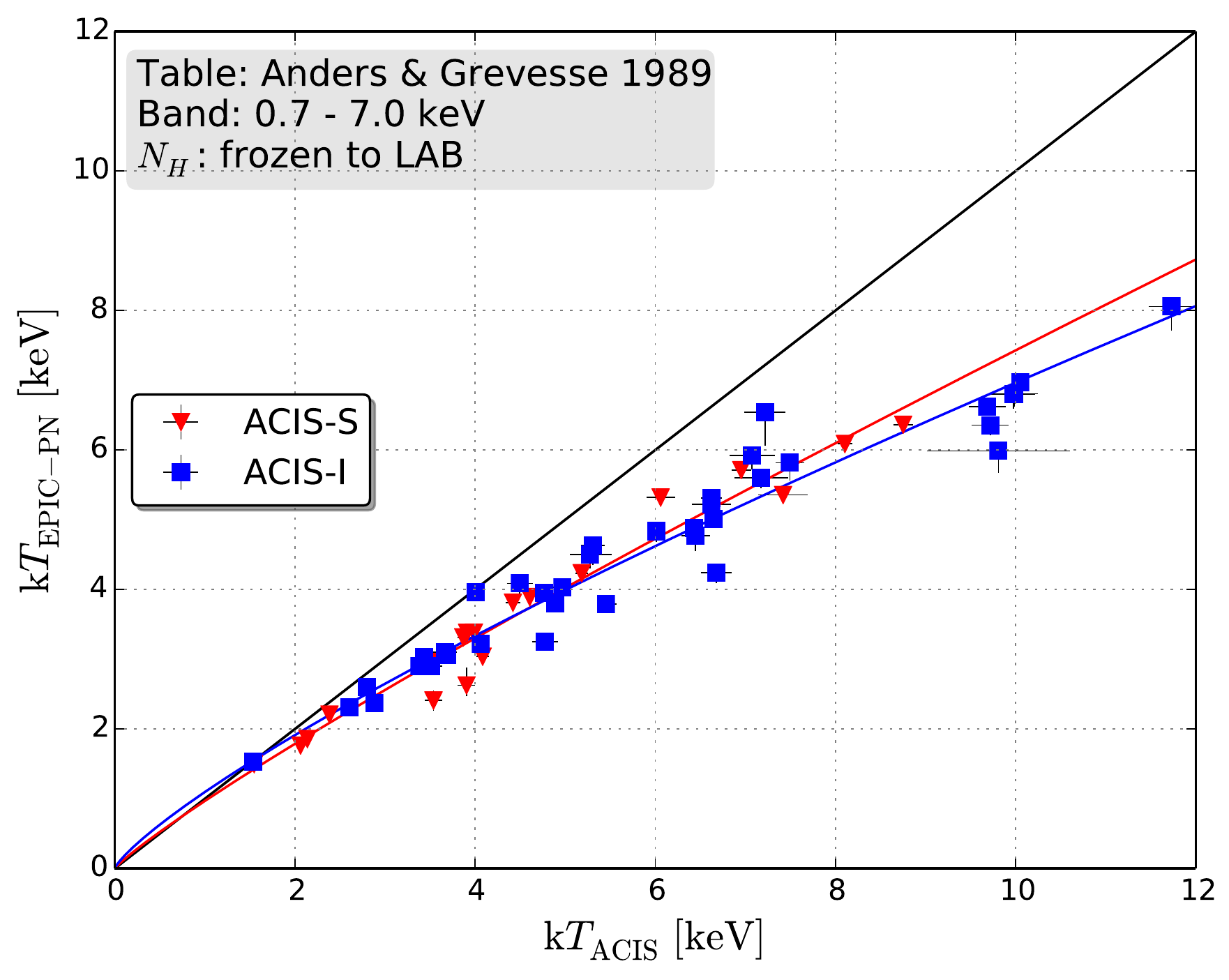}}
  \caption{Comparison of EPIC-PN full energy band temperatures with those obtained with ACIS-I (blue squares) and ACIS-S (red triangles). The $N_{\rm H}$ is frozen to the radio value of the LAB survey. For a comparison of the resulting best-fit parameters see also Fig. \ref{fig:deg2_log} and Table \ref{tab:fitparam}. The red and blue lines show the powerlaw best-fit function (Eq. \ref{eq:quadfunc}) to the ACIS-S and ACIS-I subsamples, respectively.}
  \label{fig:acisis}
\end{figure}

As indicated by the consistent ACIS-I/PN and ACIS-S/PN stacked residuals, the ACIS-I and ACIS-S temperatures are consistent in all bands (see Fig. \ref{fig:acisis} and \ref{fig:deg2_log}, right). 
 
\begin{figure}
  \resizebox{0.95\hsize}{!}{\includegraphics{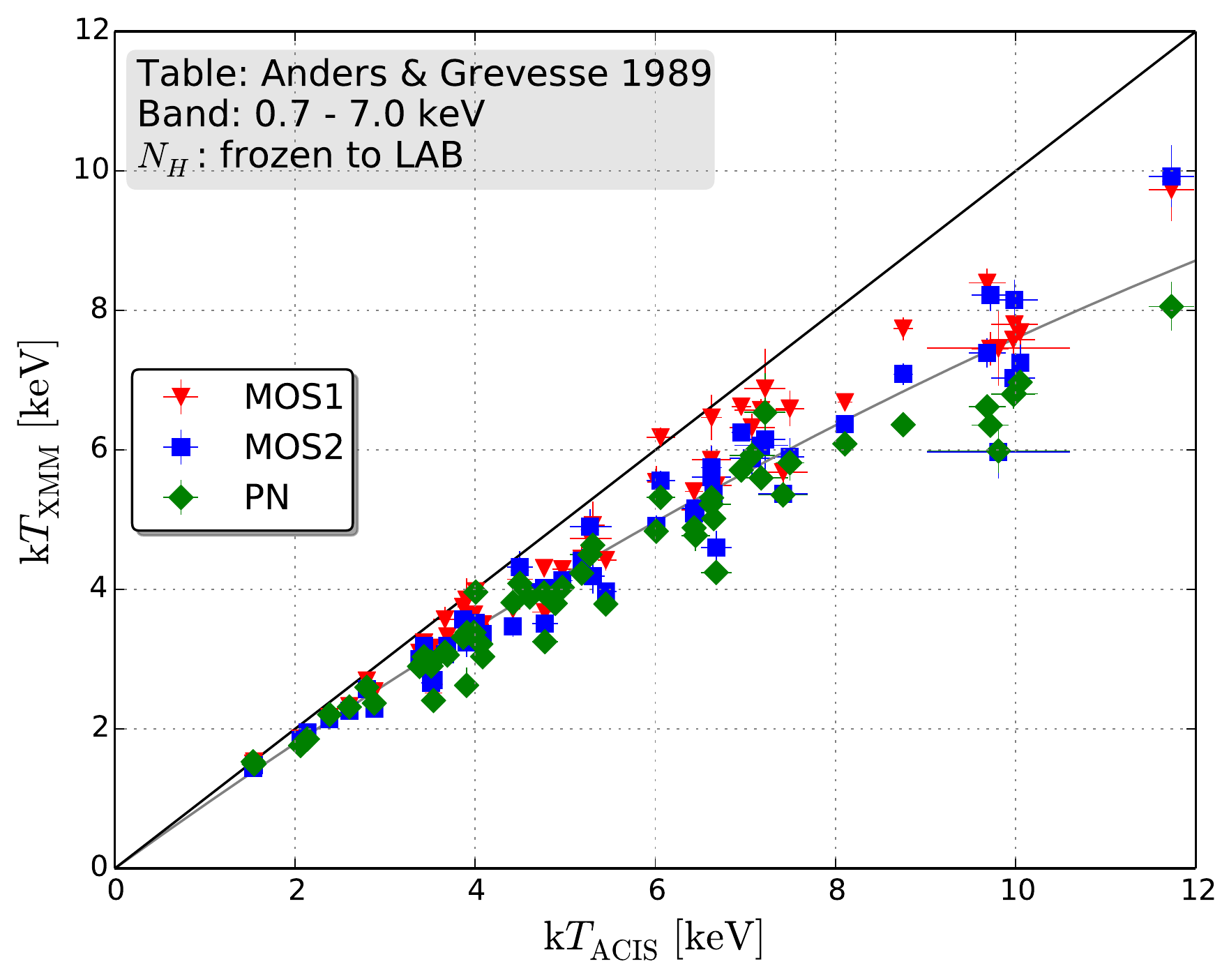}}
  \caption{Comparison of the full energy band temperatures obtained with the three individual XMM-Newton detectors (every detector combination has 56 objects) with those obtained with ACIS. The $N_{\rm H}$ is frozen to the radio value of the LAB survey. The gray line shows the powerlaw best-fit function (Eq. \ref{eq:quadfunc}) to the simultaneously fitted XMM-Newton temperatures, see also Fig.\ref{fig:deg_log_combined}.}
  \label{fig:xmm_chan}
\end{figure}

In the hard band, as indicated by the flat stack residuals, the temperatures are more consistent (most of the clusters with a $\xi < 3$, see Table \ref{tab:significance}). The feature of the MOS2/PN declining stacked residuals above $\SI{4.5}{keV}$ is not seen in the temperature comparison. This might be explained by the low statistical weight that this band gets in the spectral fit due to the low number of counts.
In the soft band, as expected due to the systematic effect in the stacked residuals, the PN temperatures are systematically lower than those of MOS1, MOS2 and ACIS, and MOS1 and MOS2 are showing very good agreement (no cluster with $\xi$ above 3).
In the full energy band the complex stacked residuals behavior as a function of energy results in MOS1 delivering higher temperatures than PN (due to soft band problems) and MOS2 delivering lower temperatures than MOS1 and ACIS and yielding approximate agreement with PN (see Figs. \ref{fig:xmm_chan}).

\begin{figure}[t]
  \resizebox{0.95\hsize}{!}{\includegraphics{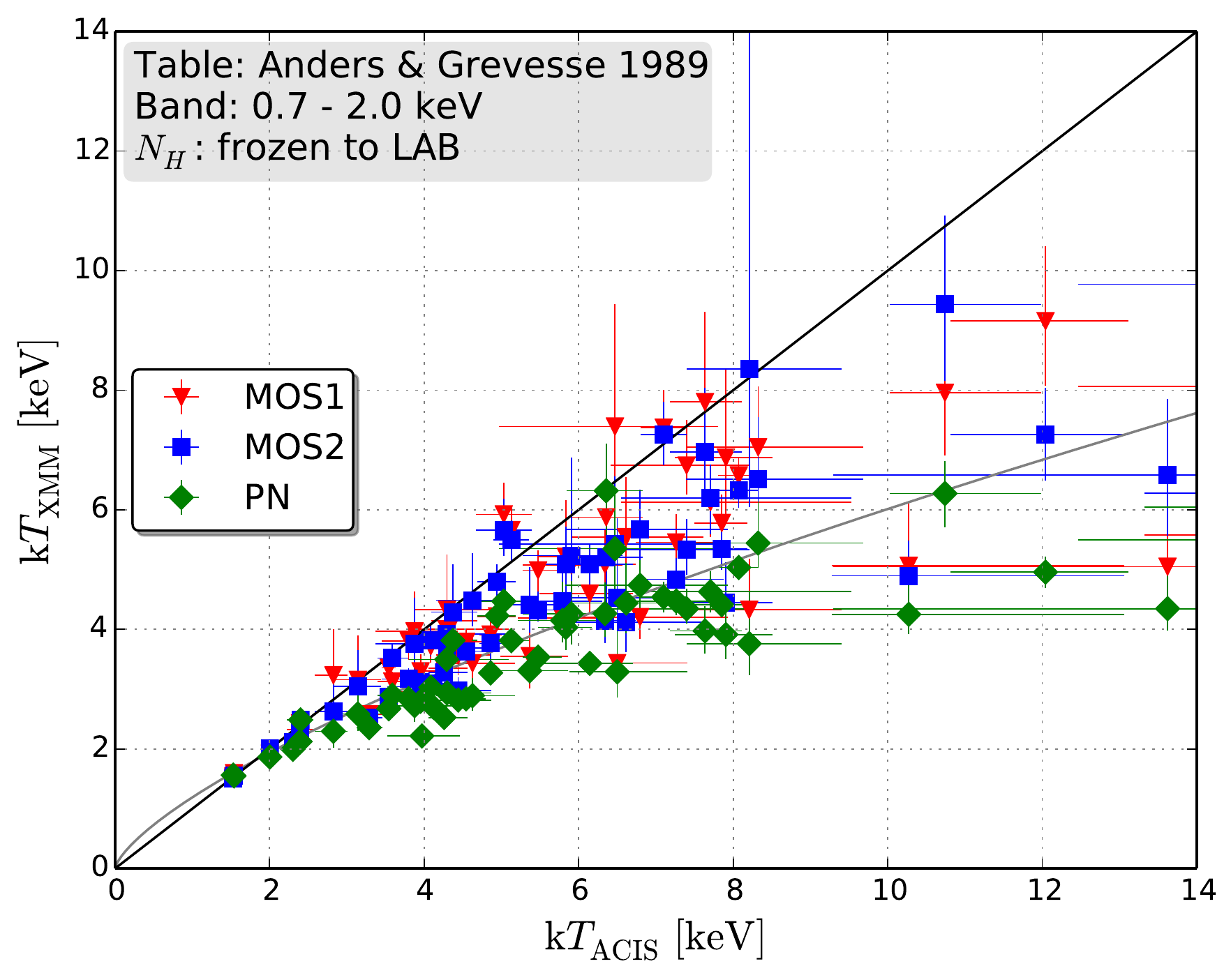}}
  \caption{Same as Fig. \ref{fig:xmm_chan}, except in the soft energy band $\SIrange{0.7}{2}{keV}$.  }
  \label{fig:xmm_chan_low}
\end{figure}

\begin{figure}[t]
  \resizebox{0.95\hsize}{!}{\includegraphics{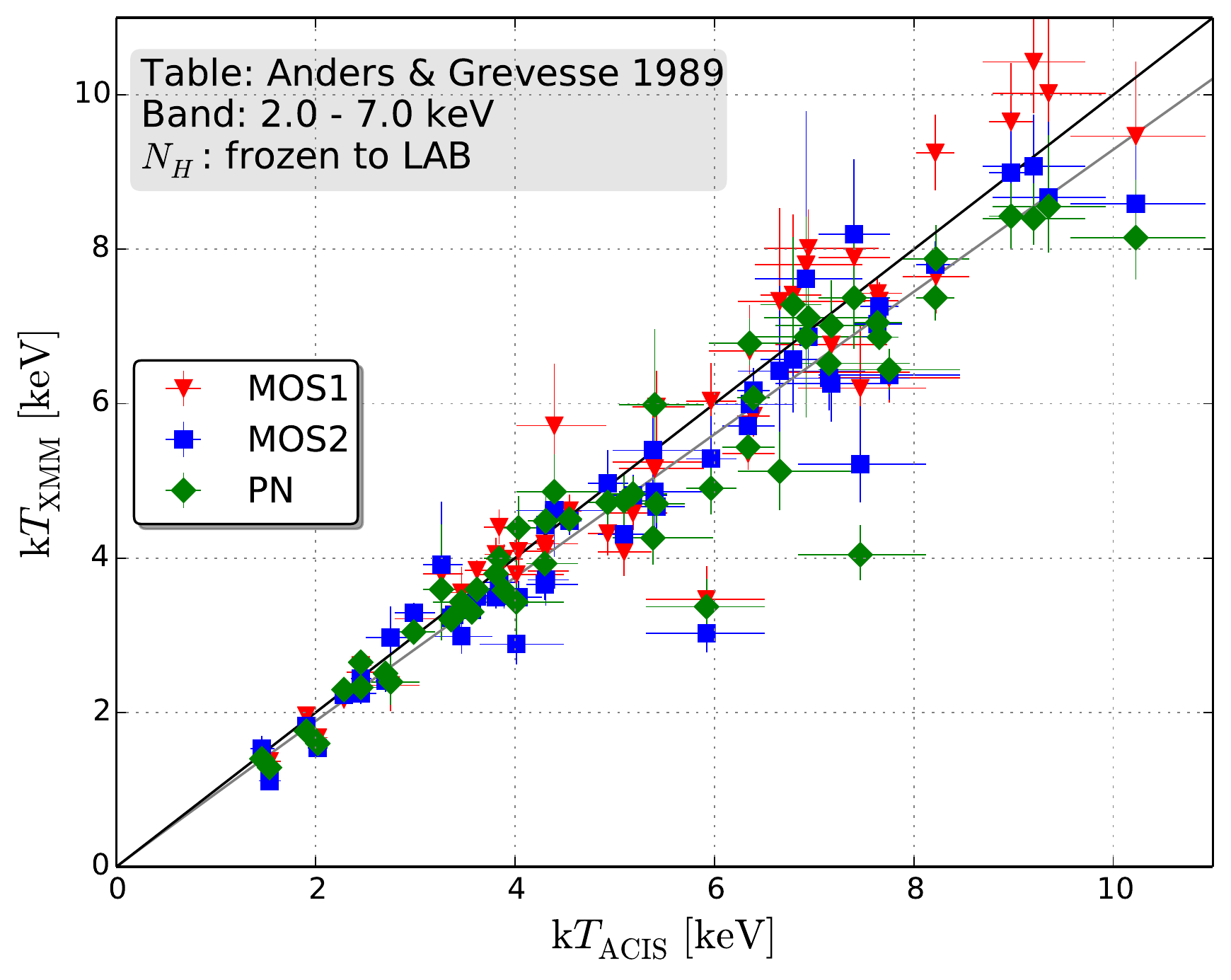}}
  \caption{Same as Fig. \ref{fig:xmm_chan}, except in the hard energy band $\SIrange{2}{7}{keV}$.}
  \label{fig:xmm_chan_high}
\end{figure}

To enable a comparison with the literature, we also combined the three XMM-Newton instruments by performing a simultaneous fit (which we will call "Combined XMM-Newton" from now on) in the different energy bands and linking temperatures and metallicities while the normalizations are free to vary. 
In the combined EPIC fit the systematic soft band stacked residuals feature for ACIS-PN  and MOS1-PN  (see Fig. \ref{fig:stacked2}) results in lower full band EPIC temperatures (see Figs.   \ref{fig:xmm_chan},  \ref{fig:xmm_chan_low}, \ref{fig:xmm_chan_high}).
The ACIS-EPIC temperature differences increase with temperature in all bands. We think this is due to the spectra of the lowest temperature clusters not having enough statistics to weight significantly the $\SIrange{1}{3}{keV}$ band cross-calibration feature.
\subsubsection{Scaling relations of temperatures between different instruments}
\begin{figure}[t]
  \resizebox{0.95\hsize}{!}{\includegraphics{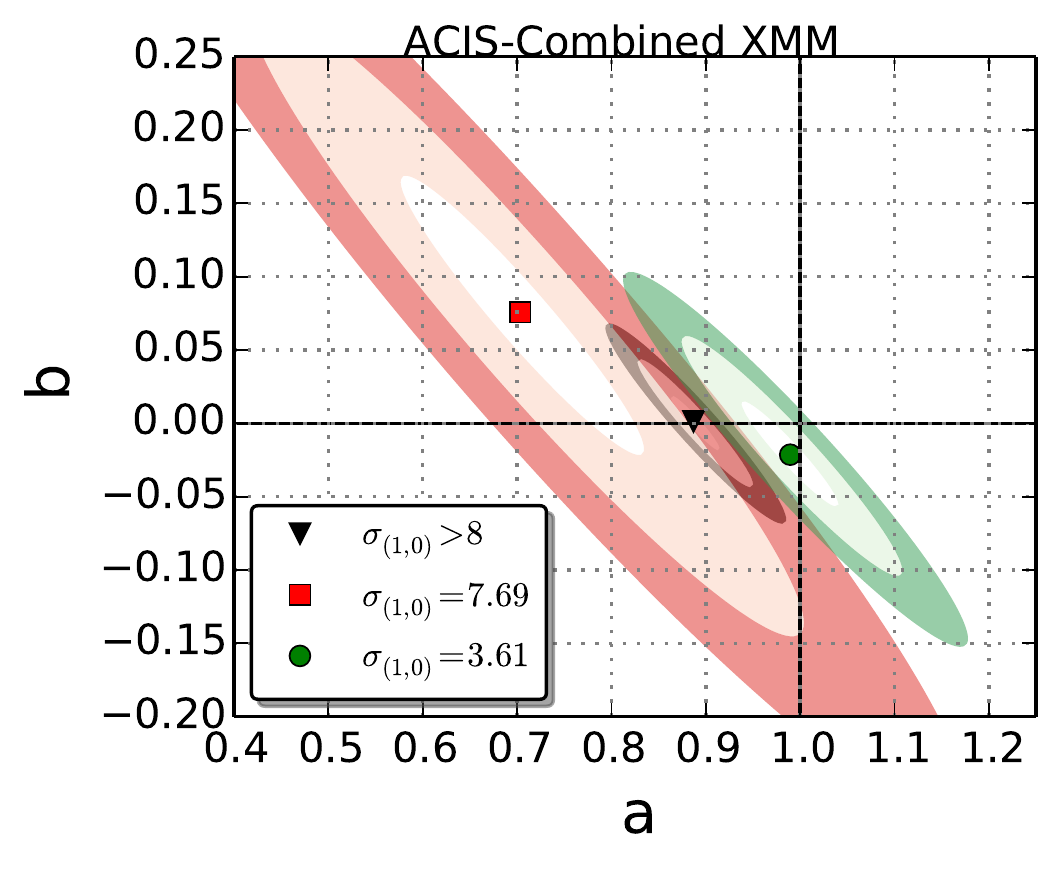}}
  \caption{Degeneracy of the fit parameters (see Eq. \ref{eq:quadfunc}) for the ACIS -- Combined XMM-Newton fit. The shaded regions correspond to the $1\sigma$, $3\sigma$ and $5\sigma$ confidence levels. The red square represents the soft energy band, the green circle the hard and the gray triangle the full band. Equality of temperatures is given for $a=1$ and $b=0$. The deviation in $\sigma$ from equality is given in the legend.}
  \label{fig:deg_log_combined}
\end{figure}

The distribution of temperature differences for any detector combination shows a Gaussian behavior in logarithmic space. We quantified this by modeling the temperatures obtained with one instrument as a powerlaw function of the values obtained with another instrument, in a given energy band, as
\begin{equation}
  \label{eq:quadfunc}
  \log_{10} \frac{kT_{\rm{I_{Y},band}}}{\SI{1}{keV}} = a \times \log_{10} \frac{kT_{\rm{I_{X},band}}}{\SI{1}{keV}} + b~.
\end{equation}
We included intrinsic scatter $\zeta$ in the fitting process (see Table \ref{tab:fitparam}), which is determined by requiring $\chi^2_{\rm red} = 1$ (like it was done e.g. by 
\citealp{2007ApJ...668..772M}). The intrinsic scatter is added in quadrature to the statistical uncertainty of the data to calculate the $\chi^2$ of the model. The degeneracy between the two parameters, $a$ and $b$, is shown in the Appendix \ref{sec:app_temp}, Figure \ref{fig:deg_log}) 
and in Figure \ref{fig:deg_log_combined} for the ACIS--XMM-Newton combined case. Neglecting the intrinsic scatter would result in tighter constraints of the fit parameters and so in higher significances of temperature differences. From Figure \ref{fig:deg_log} we conclude, that for our sample the temperatures deviate for all detector combinations at least by $5\sigma$ in the full energy band, while in the soft energy band only MOS1 and MOS2 show good agreement. In the hard band no instrument combination shows deviations larger than $\num{4.4}\sigma$. Note that for individual clusters the average significance of differences between temperatures is smaller as shown in Section \ref{ch:temps_first}.

We see more than $4\sigma$ deviation between ACIS and PN even in the hard band. Looking at the stacked residuals ratio for ACIS/PN (Fig. \ref{fig:stacked2}, top left) we see an increase in the $\SIrange{2}{3}{keV}$ band, which might be responsible for the temperature difference in the hard band. However this $4\sigma$ deviation is still small compared to the other bands.
N10 concluded that the hard band temperatures of ACIS and PN are in agreement ($\num{0.88}\sigma$ according to our analysis method), which was probably driven by the lower number of objects in the N10 sample. Despite the fact that N10 found consistent ACIS and PN temperatures in the hard band and we detect a more than $4\sigma$ deviation, the ellipses in the a-b-plane show overlap (Fig. \ref{fig:deg2_log}, right panel). Since we use more than 5 times more objects and compare here two high precision instruments, our significance increases.

In the full band, EPIC-PN temperatures are on average 29\% lower than ACIS temperatures at a cluster temperature of $\SI{10}{keV}$ (see Tab. \ref{tab:fitparam}). All detector combinations are plotted individually in Figure \ref{fig:t_all_log_full} for the full energy band, \ref{fig:t_all_log_soft} for the soft and \ref{fig:t_all_log_hard} for the hard energy band. 
Within the uncertainties, the best-fit N10 relation is in agreement with our full band relation (see Fig. \ref{fig:deg2_log}), implying persistent calibration uncertainties since 2009. Although this work deals with the Chandra CALDB 4.5.5.1, we cross checked our results of ACIS using the new ACIS QE contamination Model vN0008 included in CALDB 4.6.1 . Comparing the Chandra CALDB 4.5.5.1 (default in this work) and 4.6.1, we get  $\SI{1.9}{\percent}$ lower temperatures ($\SI{2.9}{\percent}$ scatter) using CALDB 4.6.1 for low and medium temperature clusters. For cluster temperatures above $\SI{8}{kev}$ we find $\SI{5.4}{\percent}$ ($\SI{4.8}{\percent}$ scatter) lower temperatures with CALDB 4.6.1 . In a future work we will study the effect of the new contamination model in more detail.

\begin{table*}[tbp]
  \centering
  \footnotesize
  \caption{Parameters $a$ and $b$ for the powerlaw fits of instrument X versus Y temperatures using the parameterization of Eq. \ref{eq:quadfunc}. We also give the parameters for the temperatures from N10 in a slightly different full/soft energy band. $\Delta$ is the resulting relative temperature difference between the two instruments at the temperature of instrument X.}
    \begin{tabular}[tbp]{llrrcrrr}
    \hline
    \multicolumn{1}{c}{Instruments} &  \multicolumn{1}{c}{Energy} & \multicolumn{1}{c}{a} & \multicolumn{1}{c}{b} & intr. scatter & $\Delta_{2\,{\rm keV}}$ & $\Delta_{5\,{\rm keV}}$ & $\Delta_{10\,{\rm keV}}$ \\
    \multicolumn{1}{c}{X--Y} &  \multicolumn{1}{c}{Band} &  &  & $\zeta_{\log_Y}$ &  &  &  \\
    \hline
     \parbox[0pt][1.6em][c]{0cm}{} ACIS--MOS1&Full & $0.920^{{}+0.005}_{{}-0.005}$ & $-0.001^{{}+0.004}_{{}-0.004}$ & 0.024 & 6\% & 12\% & 17\%\\
	\parbox[0pt][1.6em][c]{0cm}{} ACIS-MOS1&Soft & $0.802^{{}+0.013}_{{}-0.013}$ & $0.060^{{}+0.009}_{{}-0.009}$ & 0.052 & 0\% & 17\% & 27\%\\
	\parbox[0pt][1.6em][c]{0cm}{} ACIS-MOS1&Hard & $1.019^{{}+0.009}_{{}-0.004}$ & $-0.021^{{}+0.005}_{{}-0.005}$ & 0.024 & 4\% & 2\% & 0\%\\
     \parbox[0pt][1.6em][c]{0cm}{} ACIS--MOS2&Full & $0.909^{{}+0.005}_{{}-0.005}$ & $-0.017^{{}+0.003}_{{}-0.004}$ & 0.024 & 10\% & 17\% & 22\%\\
	\parbox[0pt][1.6em][c]{0cm}{} ACIS-MOS2&Soft & $0.789^{{}+0.009}_{{}-0.009}$ & $0.058^{{}+0.007}_{{}-0.007}$ & 0.043 & 1\% & 19\% & 30\%\\
	\parbox[0pt][1.6em][c]{0cm}{} ACIS-MOS2&Hard & $1.028^{{}+0.009}_{{}-0.009}$ & $-0.048^{{}+0.007}_{{}-0.007}$ & 0.043 & 9\% & 6\% & 5\%\\
     \parbox[0pt][1.6em][c]{0cm}{} ACIS--PN&Full & $0.836^{{}+0.005}_{{}-0.005}$ & $0.016^{{}+0.004}_{{}-0.004}$ & 0.029 & 7\% & 20\% & 29\%\\
     \parbox[0pt][1.6em][c]{0cm}{} ACIS--PN[N10]&Full & $0.837^{{}+0.007}_{{}-0.010}$ & $0.053^{{}+0.007}_{{}-0.007}$ & 0.023 & -1\% & 13\% & 22\%\\
     \parbox[0pt][1.6em][c]{0cm}{} ACIS[S]-PN&Full & $0.885^{{}+0.010}_{{}-0.010}$ & $-0.014^{{}+0.005}_{{}-0.007}$ & 0.030 & 11\% & 20\% & 26\%\\
     \parbox[0pt][1.6em][c]{0cm}{} ACIS[I]-PN&Full & $0.803^{{}+0.007}_{{}-0.007}$ & $0.040^{{}+0.005}_{{}-0.004}$ & 0.028 & 4\% & 20\% & 30\%\\
	\parbox[0pt][1.6em][c]{0cm}{} ACIS-PN&Soft & $0.652^{{}+0.009}_{{}-0.009}$ & $0.074^{{}+0.007}_{{}-0.007}$ & 0.041 & 7\% & 32\% & 47\%\\
     \parbox[0pt][1.6em][c]{0cm}{} ACIS-PN[N10]&Soft & $0.769^{{}+0.017}_{{}-0.020}$ & $0.075^{{}+0.014}_{{}-0.014}$ & 0.038 & -1\% & 18\% & 30\%\\
	\parbox[0pt][1.6em][c]{0cm}{} ACIS-PN&Hard & $0.947^{{}+0.009}_{{}-0.009}$ & $0.006^{{}+0.007}_{{}-0.005}$ & 0.034 & 2\% & 7\% & 10\%\\
     \parbox[0pt][1.6em][c]{0cm}{} ACIS-PN[N10]&Hard & $0.926^{{}+0.010}_{{}-0.014}$ & $0.058^{{}+0.009}_{{}-0.009}$ & 0.020 & -9\% & -2\% & 4\%\\
     \parbox[0pt][1.6em][c]{0cm}{} ACIS--Combined XMM&Full & $0.889^{{}+0.005}_{{}-0.003}$ & $0.000^{{}+0.004}_{{}-0.004}$ & 0.025 & 7\% & 16\% & 23\%\\
	\parbox[0pt][1.6em][c]{0cm}{} ACIS-Combined XMM&Soft & $0.703^{{}+0.021}_{{}-0.021}$ & $0.076^{{}+0.018}_{{}-0.016}$ & 0.129 & 3\% & 26\% & 40\%\\
	\parbox[0pt][1.6em][c]{0cm}{} ACIS-Combined XMM&Hard & $0.989^{{}+0.009}_{{}-0.009}$ & $-0.021^{{}+0.005}_{{}-0.007}$ & 0.042 & 5\% & 6\% & 7\%\\
     \parbox[0pt][1.6em][c]{0cm}{} MOS1--MOS2&Full & $0.983^{{}+0.003}_{{}-0.003}$ & $-0.012^{{}+0.002}_{{}-0.003}$ & 0.014 & 4\% & 5\% & 6\%\\
     \parbox[0pt][1.6em][c]{0cm}{} MOS1--MOS2&Soft & $0.976^{{}+0.009}_{{}-0.010}$ & $0.006^{{}+0.005}_{{}-0.005}$ & 0.008 & 0\% & 3\% & 4\%\\
     \parbox[0pt][1.6em][c]{0cm}{} MOS1--MOS2&Hard & $1.001^{{}+0.010}_{{}-0.009}$ & $-0.021^{{}+0.006}_{{}-0.006}$ & 0.030 & 5\% & 4\% & 4\%\\
     \parbox[0pt][1.6em][c]{0cm}{} MOS1--PN&Full & $0.908^{{}+0.003}_{{}-0.003}$ & $0.019^{{}+0.003}_{{}-0.002}$ & 0.013 & 2\% & 10\% & 16\%\\
     \parbox[0pt][1.6em][c]{0cm}{} MOS1--PN&Soft & $0.785^{{}+0.010}_{{}-0.010}$ & $0.040^{{}+0.006}_{{}-0.005}$ & 0.027 & 6\% & 22\% & 33\%\\
     \parbox[0pt][1.6em][c]{0cm}{} MOS1--PN&Hard & $0.947^{{}+0.007}_{{}-0.007}$ & $0.016^{{}+0.005}_{{}-0.005}$ & 0.014 & 0\% & 5\% & 8\%\\
     \parbox[0pt][1.6em][c]{0cm}{} MOS2--PN&Full & $0.921^{{}+0.005}_{{}-0.003}$ & $0.031^{{}+0.003}_{{}-0.003}$ & 0.017 & -2\% & 5\% & 10\%\\
     \parbox[0pt][1.6em][c]{0cm}{} MOS2--PN&Soft & $0.802^{{}+0.009}_{{}-0.009}$ & $0.038^{{}+0.005}_{{}-0.005}$ & 0.018 & 5\% & 21\% & 31\%\\
     \parbox[0pt][1.6em][c]{0cm}{} MOS2--PN&Hard & $0.940^{{}+0.007}_{{}-0.007}$ & $0.038^{{}+0.005}_{{}-0.005}$ & 0.018 & -5\% & 1\% & 5\%\\
  \end{tabular}
  \label{tab:fitparam}
\end{table*}

We want to mention here that we also tested the self consistency of the instruments by comparing the soft and hard band temperatures of the same instrument. However, the conclusions of these results depend strongly on the multi temperature structure of the ICM. Details are provided in Appendix \ref{ch:self-consistency}.

\begin{figure}[t]
  \resizebox{0.95\hsize}{!}{\includegraphics{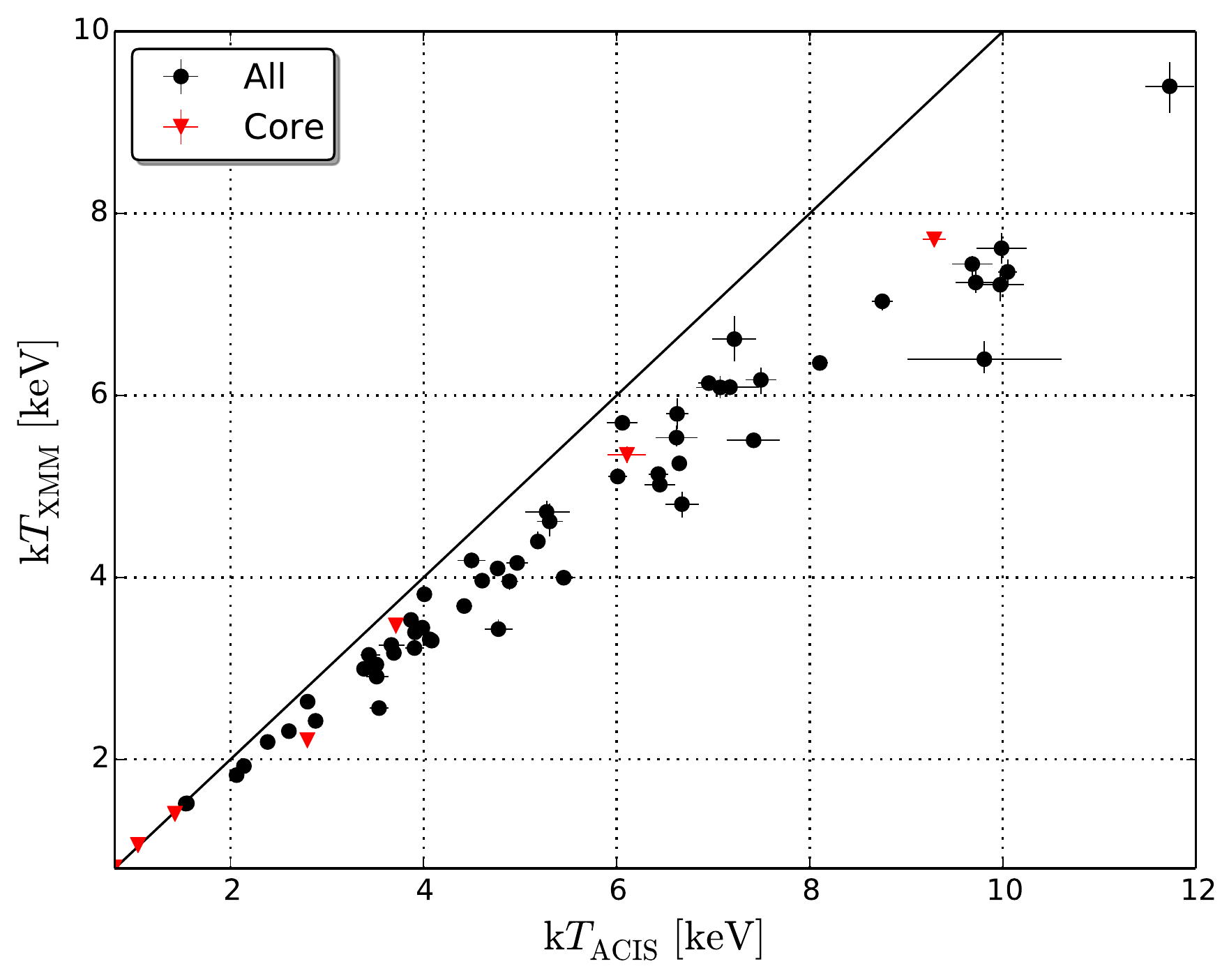}}
  \caption{Combined XMM-Newton versus Chandra temperature. The excluded objects (too large core radius) are marked here by red triangles.}
  \label{fig:excluded}
\end{figure}

The combined EPIC and ACIS temperatures indicate that the clusters that are excluded because of the too large cool core radius (see Section \ref{sec:properties}) do not show a systematic behavior compared to the other clusters (see Fig. \ref{fig:excluded}). 

\begin{figure}
  \resizebox{\hsize}{!}{\includegraphics{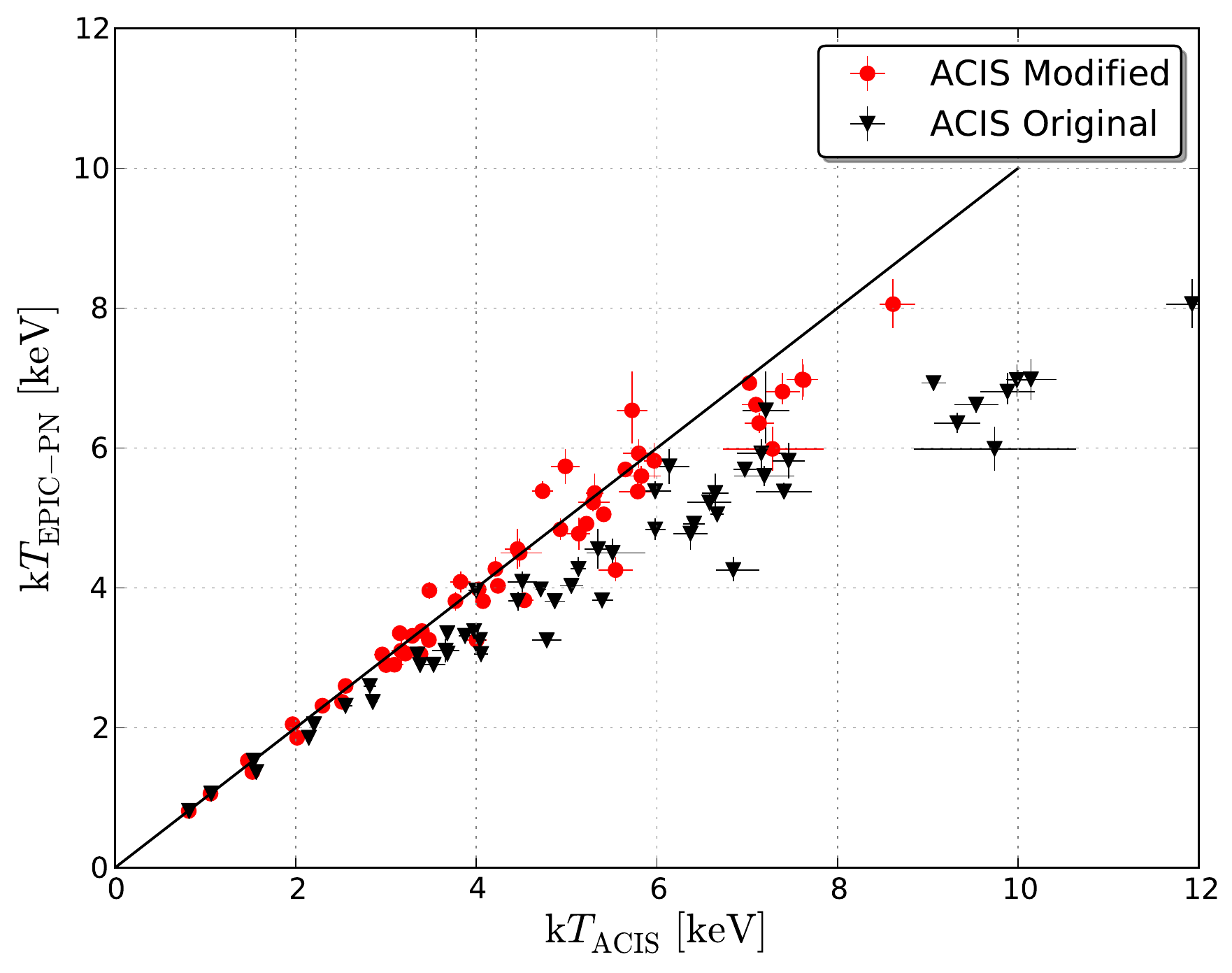}}
  \caption{Temperatures before and after modifying the ARF files of ACIS spectra according to the energy-dependent calibration uncertainties shown in Fig. \ref{fig:stacked2}. Note that this plot shows the 53 clusters from the stacked residuals ratio sample (Section \ref{ch:srr}).}
  \label{fig:modified_arf}
\end{figure}

We modeled the energy dependence of the cross-calibration uncertainties by cubic spline interpolation (e.g. \citealp[Chapter 3.3]{1992nrca.book.....P}). This was applied in Fig. \ref{fig:stacked}, \ref{fig:stacked2} and Table \ref{tab:spline} to the stacked residuals ratios (see Section \ref{ch:stackedresiduals}).
These splines can be used to estimate the effect of the effective area cross-calibration uncertainties on the spectral analysis performed with a given EPIC or ACIS instrument. The effective area of a given instrument must be multiplied by the corresponding spline to obtain best-fit temperatures assuming that the reference instrument is precisely calibrated.
To convert e.g. from ACIS to EPIC-PN, one should multiply the effective area of ACIS with the energy-dependent spline values of ACIS/PN from Table \ref{tab:spline}. 
The underlying assumption is here, that the reason for the temperature differences between the detectors is the uncertainty in the effective area calibration.
As an example we pick the most extreme case, ACIS versus EPIC-PN: We find that indeed temperatures between ACIS and EPIC-PN (Fig. \ref{fig:modified_arf}) are now consistent. 
The significance of temperature differences between ACIS and PN is now $2.7\,\sigma$ in the full band, while before we had a more than $8\,\sigma$ deviation.
To enable the use of our results for calibration work, we present the spline parameters for each instrument pair in Table \ref{tab:spline} and we provide a tool to modify the effective area based on the splines of the stacked residuals ratios\footnote{\url{https://wikis.mit.edu/confluence/display/iachec/Data3}}.

\section{Discussion}\label{sec:disc}
In the previous section we have shown that there exist systematic temperature differences between ACIS and the EPIC detectors. We investigate the consequences arising when our assumption of a single temperature plasma breaks down and we have to deal with a multi temperature structure. Finally, we give a comparison of measured and independently derived hydrogen column density values and estimate the cosmological impacts of our results.
\subsection{Systematics}\label{sec:systematics}
\subsubsection{Multiphase ICM}\label{sec:multi_icm}
While we know the relative calibration uncertainties from the stacked residuals ratios (see Section \ref{ch:stackedresiduals}) we discuss here the possibility that the observed temperature differences between the instruments and detectors are not caused by calibration uncertainties but by the different effective areas and a multi temperature structure of the ICM. The role of the multi phase ICM has been discussed in detail e.g. in \cite{2004MNRAS.354...10M,2006ApJ...640..710V,2013SSRv..177..195R}.
As shown in \cite{2013SSRv..177..195R}, one can conclude from the effective areas of the instruments that Chandra is more sensitive to the harder spectra and higher temperatures than XMM-Newton, which may explain the higher temperatures if multi temperature structure plays an important role. In the following we will quantify if this effect is significant and a multi temperature structure is important by performing simulations. We show later that the restrictions on the two temperature plasma cannot be fulfilled. We will also show examples, where we expect a strong multi phase structure of the ICM like the cool core regions of CC clusters or the full ($\SI{3.5}{\arcmin}$) region of clusters with a cool core larger than this size. Finally we compare the results from Chandra events files which were smoothed to the XMM-Newton resolution.

In summary, we demonstrate that none of the studied possible effects can explain the observed temperature differences.

\paragraph{Simulations}\label{ch:icm_simu}
To test the effect on the measurements of a multi temperature plasma we 
simulated a set of spectra for the different XMM-Newton detectors and for 
ACIS-I following the same approach presented in \citet[Sect. 4.6]{2013SSRv..177..195R}. 
To isolate the influence of multi temperature structure on best-fit temperatures we assume here that the instruments are perfectly calibrated.
We used as test cases for the cold component three different temperatures: 0.5, 1, and $\SI{2}{keV}$; and 
three different metal abundances (0.3, 0.5 and 1 times solar). The 
temperature of the hot component was varied from 3 to $\SI{10}{keV}$ while the 
metallicity was kept fixed to 0.3 solar, a value typical for the clusters in our sample. The relative contribution of 
the second plasma component has been estimated by varying the  emission 
measure ratios (from $10\%$ to equal emission measure).
We then fitted the resultant spectra with a single temperature model 
and also left the metallicity free to vary. The best-fit values and 
the 68\% errors have been taken from the distribution of 100 
realizations. 

\begin{figure}
  \resizebox{\hsize}{!}{\includegraphics{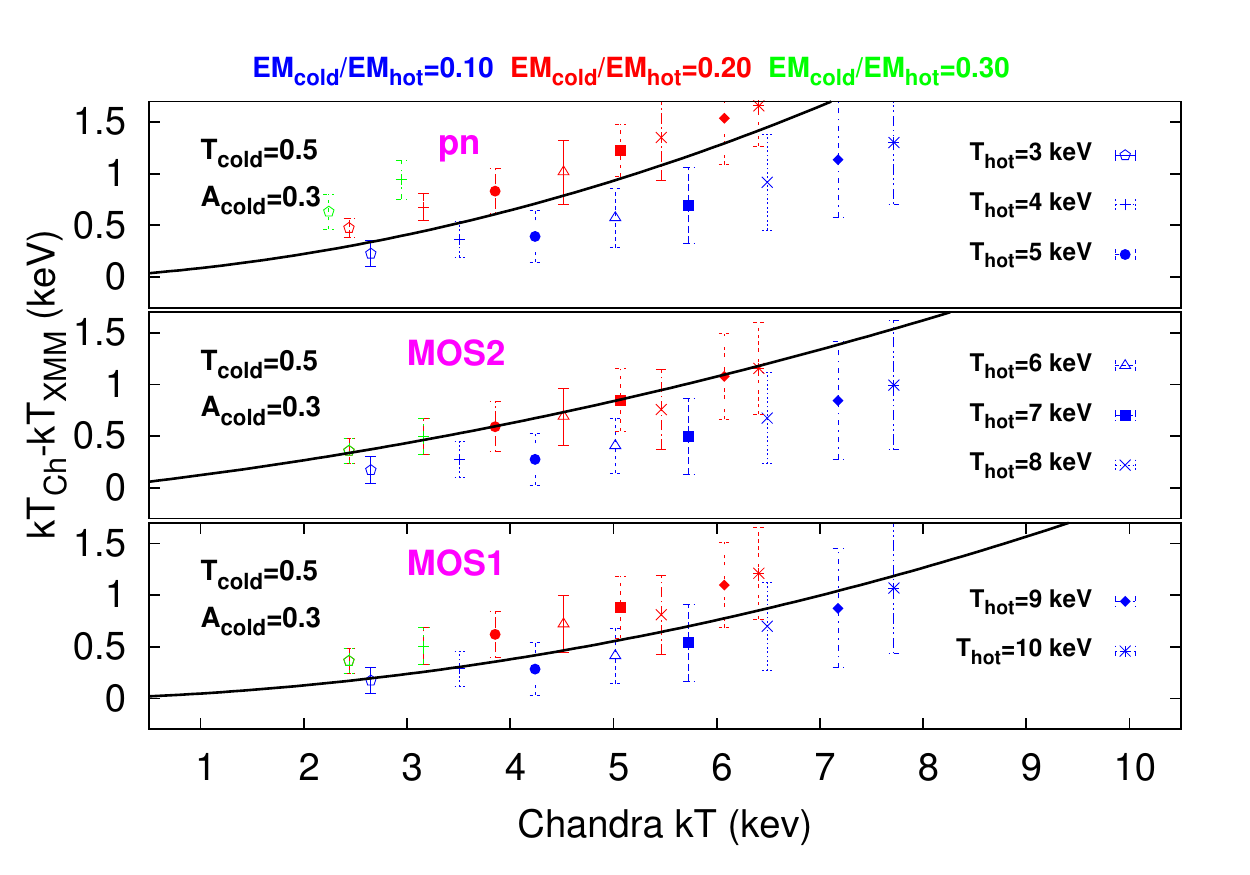}}
  \caption{The difference between Chandra and XMM-Newton single temperature fits to two-temperature component mock-spectra versus the Chandra best-fit temperature. The three colors correspond to different emission measure ratios (EMR) of the cold and hot component, the symbols represent 8 different temperatures of the hot component, while its metallicity is frozen to 0.3 of the solar value and the redshift to 0.05, which corresponds to the mean HIFLUGCS redshift. Temperature and metallicity of the cold component are always fixed to 0.5 and 0.3, respectively. The black curve shows the measured temperature difference between Chandra and XMM-Newton as presented in Table \ref{tab:fitparam}.}
  \label{fig:det_comp}
\end{figure}

We found that when the cold component is $\sim\SI{2}{keV}$ the fit is always 
good ($\chi^2 <$ 1.1) independent of the amount of "cold gas" present in the cluster 
and of the metallicity of this gas. For cooler temperatures the amount 
of cold gas and its metallicity play an important role. Clearly, if the 
cold gas would be very enriched the fit would yield a bad ($\chi^2 >$ 2) in almost all the 
cases (i.e. also with only $10\%$ of cold gas with respect to the hot 
component) while for lower abundance it depends also on the amount of 
gas. Since in the observations we do not see such bad $\chi_{\rm red}^2$, we excluded 
all the combinations with a $\chi_{\rm red}^2 > 2$.
Each dataset of the simulated spectra contains $\num{10000}$ counts. This is a conservative approach, since we typically have more counts in our spectra from the observed data. Due to the higher statistics of the observed spectra with respect to the simulated ones, we would find an inappropriate model (e.g. due to multi temperature structure) first in the observed data.
 We checked  the difference in 
temperature between Chandra and XMM-Newton of the simulated spectra and we found that only when the 
second component is very cold (i.e. $\SI{0.5}{keV}$) the fitted temperature of 
the two instruments starts to be significantly different. In Fig. \ref{fig:det_comp} we 
show only this case.
We note that to explain the observed difference between Chandra and EPIC-MOS2
an emission measure ratio (EMR) of almost 0.2 is required.  We will show in the next 
paragraphs that this is not the case.
Furthermore, an ${\rm EMR}=0.2$ is not able to explain the observed difference 
between Chandra and MOS1/PN. Thus,  we conclude  that although a multi 
temperature ICM cannot be excluded, it cannot explain 
the observed difference in temperature between the different instruments. We also note here that a higher EMR gives always a bad $\chi^2$ that we do not observe in the real observations.

\paragraph{Two Temperature fits}
To test the results from the simulations we selected the clusters with the highest number of counts in both, XMM-Newton and Chandra: 
Abell 2029 has $\num{210000}$ counts in Chandra and roughly $\num{62000}$ counts in PN and a Chandra temperature of $\SI{8.7}{keV}$. For both instruments we fitted a two temperature model, fixing the temperature of the second component to $\SI{0.5}{keV}$ and the metallicity to $\num{0.3}$. The temperature of the first component increased for both instruments by $3-4\%$ compared to the single temperature model. The normalization of the second component is for both instruments $1-5\%$ of the first component. This is far too low to explain any temperature differences obtained by the two instruments. Also fixing the normalization of the second component to $20\%$ of the first component increases the reduced $\chi^2$ to almost 4 for PN and $>15$ for ACIS. 

Another good example is the cool core cluster Abell 2142 with more than $\num{180000}$ counts in both instruments and a Chandra temperature above $\SI{9}{keV}$. This cluster has a cool core radius larger than $\SI{3.5}{\arcmin}$, so in an observation within this threshold more than one temperature component should be present. For Chandra the normalization of a $\SI{0.5}{keV}$ component is consistent with zero. This is also true if the temperature of the cold component is frozen to $\SI{2}{keV}$. For PN the normalization of the $\SI{0.5}{keV}$ component is below $1.5\%$ and for a fitted temperature of $\SI{2.7}{keV}$ of the cold component the normalization is about $17\%$ of the first component. From the simulations we conclude that every cold component above $\SI{2}{keV}$ cannot explain the observed difference, so we cannot find any hint for a multi phase ICM being the reason for the inconsistency of the instruments.

\begin{figure}
  \resizebox{\hsize}{!}{\includegraphics{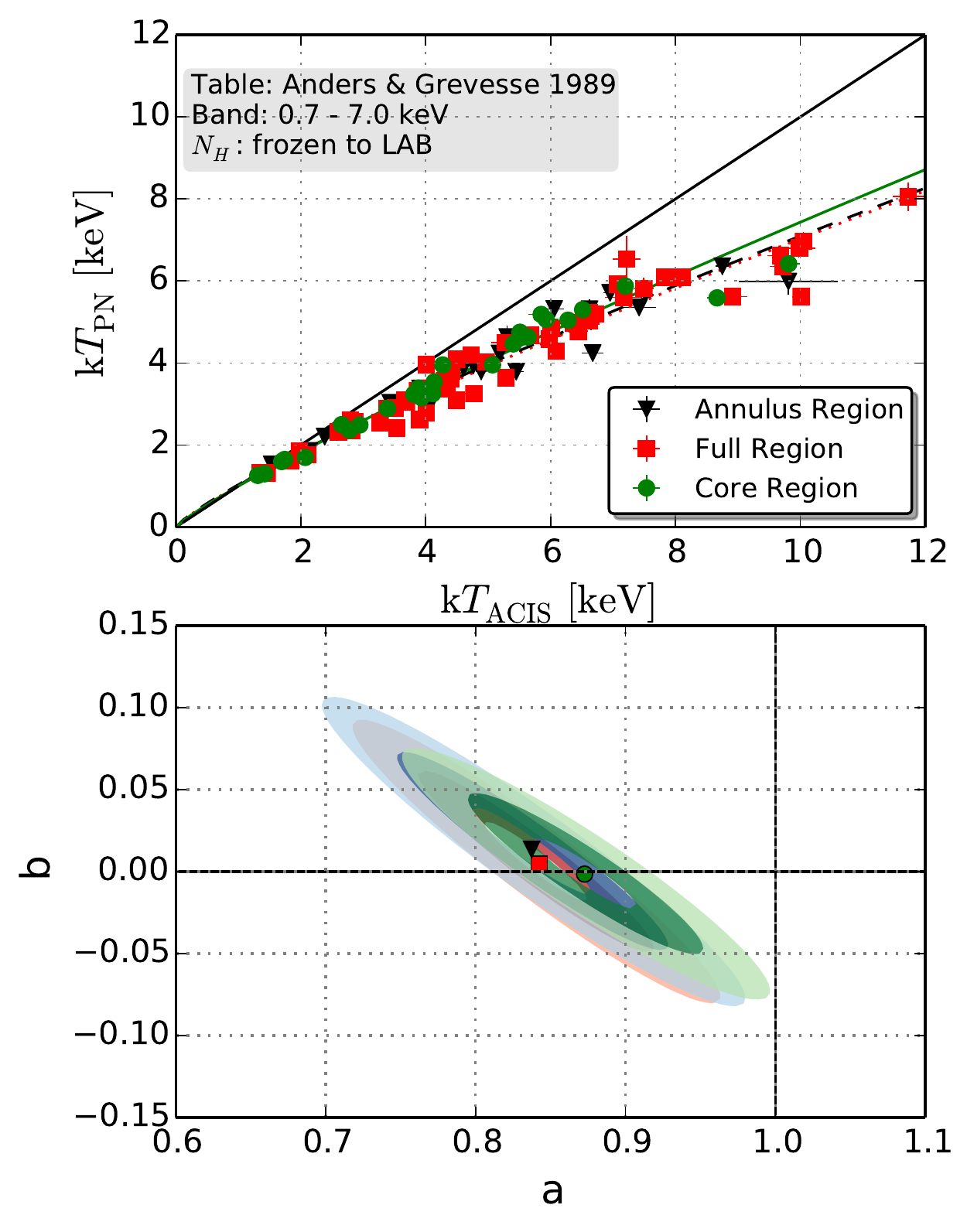}}
  \caption{\textit{Top: }XMM-Newton EPIC-PN versus Chandra ACIS temperatures in the full energy band. The different colors correspond to the 3 different regions, where the spectra could be extracted: The full $\SI{3.5}{\arcmin}$ circle (red squares), the cool-core region (green circles) and the annulus between the two (black triangles). In the cool core region, the strongest influence of multi temperature plasma is expected. \textit{Bottom: }Degeneracy between the parameters $a$ and $b$ for the three cases shown above. The shaded regions correspond to $1\sigma$, $3\sigma$ and $5\sigma$ confidence level.}
  \label{fig:core_full}
\end{figure}

\paragraph{Multitemperature plasmas}
Since galaxy clusters in general, and especially cool core clusters, do not show only one dominant temperature, but a temperature profile, we select now regions, where a clear multi phase ICM structure is present: The cool cores of CC clusters and the whole circular region until $\SI{3.5}{\arcmin}$. 
Assuming multi temperature structure to contribute significantly to the observed temperature differences, we would then expect to find significantly larger temperature differences. We do not observe this, though.
We have already demonstrated in Figure \ref{fig:excluded} that the restriction to isothermal regions does not introduce any bias, but especially for the cool core regions we verify this again. We ended up with 28 cool core region clusters, 56 clusters where we used the full circular region and compared this to the 28 clusters using the annulus region. To choose the extreme case we only use the PN detector for XMM-Newton. As it can be seen in Fig. \ref{fig:core_full} the best-fit curves do agree and we can be sure that multi temperature structure of clusters has no significant effect on our measurements of the instrumental differences.
 
\paragraph{Summary of the Multiphase ICM tests}
In this Section we have shown that assuming the observed temperature differences are only caused by multi temperature structure of the ICM, i.e. the calibrations are perfect, we can set up restrictions on the cold plasma component (${\rm EMR}\approx 20\%$ and ${\rm kT} = \SI{0.5}{keV}$). These restrictions were derived by fitting two simulated plasma components with a single component and requiring the observed temperature difference between detectors to be reproduced with the simulations as well as $\chi_{\rm red}^2 < 2$. Applying this to real high quality observations by fitting two plasma components we cannot recover the required emission measure of the cold component. Also we cannot see any hint of a different behavior of the temperature difference for regions with clear multi phase structure (e.g. the cool core region), which would be expected, if the multi temperature structure is responsible for the measured temperature differences.

\subsubsection{Temperature - Abundance Degeneracy}\label{ch:tempabun}
The two parameters, temperature and abundance of heavy elements, are not completely independent of each other (e.g. \citealp{2010A&A...522A..34G,2000MNRAS.311..176B}). Since for many of our clusters the constrained abundance is also not in agreement within the different instruments, it is possible that a degeneracy of the two parameters is producing inconsistent values for both of them. To make sure that this degeneracy is not the reason of the detected temperature difference, we refitted the XMM-Newton data of all clusters freezing the metallicity to the one obtained by Chandra. On average the EPIC-PN temperature increased by 0.8\% compared to the best-fit temperature where the relative abundance is free to vary. For only 8 of the 56 clusters the EPIC-PN temperature increased by more than 2\% (at maximum by 8.7\%), but for none of these cases the new (with frozen relative abundance to Chandra ACIS) EPIC-PN temperature is in agreement with Chandra ACIS.
Therefore, any inconsistency of the temperatures cannot be explained by wrongly constrained abundance, also because the temperature change described in this Section is not systematic in one direction.

\subsubsection{Effects of the Point Spread Function (PSF)}
One last case to discuss here is the possible scatter of the cool core emission into the annular region due to the broader XMM-Newton PSF. We used the Chandra cleaned events files and redistributed the position of all photons using a Gaussian smoothing kernel with a $\sigma$ between $\SI{1}{\arcsec}$ and $\SI{70}{\arcsec}$. This scenario is different than what we simulated in Section \ref{ch:icm_simu} since we assumed there that the detectors ``see'' the same photon distribution, but are sensitive to different parts of the spectrum, which is more complicated than our one-component model. 
Now the instrument has to deal with different photon distributions caused by the different PSFs. We tested this for four clusters, where we detect a temperature difference between XMM-Newton and Chandra and which have a strong cool core: Abell 85, Abell 133, Abell 2204 and Abell 3112. In all four cases the temperature is very constant up to a threshold of the smoothing kernel $\sigma$. A significant drop of the temperature is detected beyond $\SI{30}{\arcsec}$, then the temperature drops constantly with increasing $\sigma$. Even if the Gaussian shape of the PSF (constant with photon energy) is not the best approximation of the real PSF behavior, we can conclude from this test that the PSF of XMM-Newton (half power diameter of $\SI{15}{\arcsec}$ on axis) is not sufficient to scatter enough emission from cooler regions. 

\subsection{Influence of $N_\mathrm{H}$ and the abundance table}\label{sec:nhtot}
In the following Section we show the influence of the heavy elements abundance table and the hydrogen column density in the absorption model. We demonstrate that the $N_{\rm H}$ constrained in a spectral fit can be compared to independently determined reference values. This comparison could help to evaluate the calibration status of a detector through the level of agreement with reverence values, but the conclusions depend on the abundance table.

The relative abundance table gives the number densities of atoms of a certain element relative to hydrogen. Most of the abundance tables are established using measurements of the solar photosphere and corona and from meteorites. It is still very common to use the (\verb+xspec+ default) table presented in \cite{1989GeCoA..53..197A}, but we also want to test for a dependence of our results on the choice of the table by using the \cite{2009ARA&A..47..481A} relative abundances, which represents more recent measurements. 
In our analysis we are able to see a systematic effect by using this latter abundance table: Temperatures derived using Aspl are on average 5\% higher. These higher temperatures are produced independent of the detector (ACIS and EPIC). Since this effect is much bigger than expected, we tested, if the absorption or the emission component is responsible for this change. Using the \verb+wabs+ instead of the \verb+phabs+-model for the absorption fixes the abundance table for the absorption model to the hard-wired abundance table from \cite{1982GeCoA..46.2363A}, while \verb|phabs| self-consistently uses the same abundance table for absorption and emission. With the \verb|wabs| model the temperature change is then very tiny and the results are always consistent within the errorbars. So the reason for this large change in temperature by using the Aspl abundance table is given in the absorption. 

The step during the spectral fitting process of accounting for the absorption of heavy elements along the line of sight is important. The absolute abundance of these metals can be traced by measuring the neutral hydrogen column density (e.g.\ with HI 21cm radio measurements) and assuming relative abundances as in the solar system. A widely used HI-survey is the Leiden/Argentine/Bonn (LAB) survey (\citealp{2005A&A...440..775K}). 
The cross-calibration uncertainties of the effective area are larger
at the lowest energies (Section \ref{ch:temps}), where the Galactic absorption is
significant. If a) one detector had its effective area very accurately
calibrated in the full energy band, and b) the emission modeling was
accurate (i.e. treating correctly the possible multi temperature structure),
the best-fit hydrogen column densities obtained with this instrument
should be consistent with the Galactic values. Thus the free $N_{\rm H}$ test could
yield some information on the calibration accuracy.

We re-fitted the full energy band EPIC-PN and ACIS spectra, allowing the temperature, metal abundance, emission measure and the hydrogen column density to be free parameters.
Using the PN detector as representative for XMM-Newton, we can conclude that Chandra finds systematically (on average around 40\%) higher $N_{\rm H}$ values than XMM-Newton. Also compared to the LAB values, a Chandra detected $N_{\rm H}$ is systematically higher (on average 20\% using AnGr), while PN detects 33\% lower column densities using AnGr.
Comparing the detected $N_{\rm H}$ values using the Aspl abundance table with the detected column densities using AnGr reveals a strong trend towards detecting $20-30$\% higher column densities with the more recent abundance table.

The dominant element responsible for this temperature increase is oxygen, since its relative abundance value is reduced by almost $50\%$ in the Aspl table compared to the default AnGr. This decreased relative oxygen abundance causes the detected $N_{\rm H}$ values to increase, since the absolute number of oxygen atoms along the line of sight should stay constant. Conversely if the $N_{\rm H}$ value is frozen, the absolute number of oxygen atoms along the line of sight is reduced by switching to the Aspl table, so the effect of absorption is lower and the temperature needs to increase to compensate for this effect.

The radio surveys measure the hydrogen column density from the HI-$\SI{21}{cm}$ line. This only provides the neutral hydrogen along the line of sight, the molecular and ionized hydrogen is not accounted for. Even if the neutral hydrogen should usually contribute most to the total hydrogen, there exists an easy method to account for the molecular hydrogen (for more details see \citealp{2013MNRAS.tmp..859W}): One can use the dust extinction $E(B-V)$ measured in the B and V band as a tracer for the molecular hydrogen. 

So we define the total hydrogen column density as
\begin{equation}
  \label{eq:nhtot}
  N_{\rm H, tot} = N_{\rm HI} + 2 N_{\rm H_2} = N_{\rm HI} + 2 N_{\rm H_2, max} \times \left( 1 - e^{-\frac{N_{\rm HI}E(B-V)}{N_c}} \right)^\alpha\,,
\end{equation}
with $N_{\rm H_2, max} = \SI{7.2e20}{cm^{-2}}$, $N_c = \SI{3e20}{cm^{-2}}$ and $\alpha = 1.1$ as given by \cite{2013MNRAS.tmp..859W}. The relation has been calibrated using X-ray afterglows of gamma ray bursts. A webtool\footnote{\url{http://www.swift.ac.uk/analysis/nhtot/index.php}} is available to calculate $N_{\rm H, tot}$ for a given position using the infrared data from IRAS and COBE/DIRBE (\citealp{1998ApJ...500..525S}). 

\begin{figure}
  \resizebox{\hsize}{!}{\includegraphics{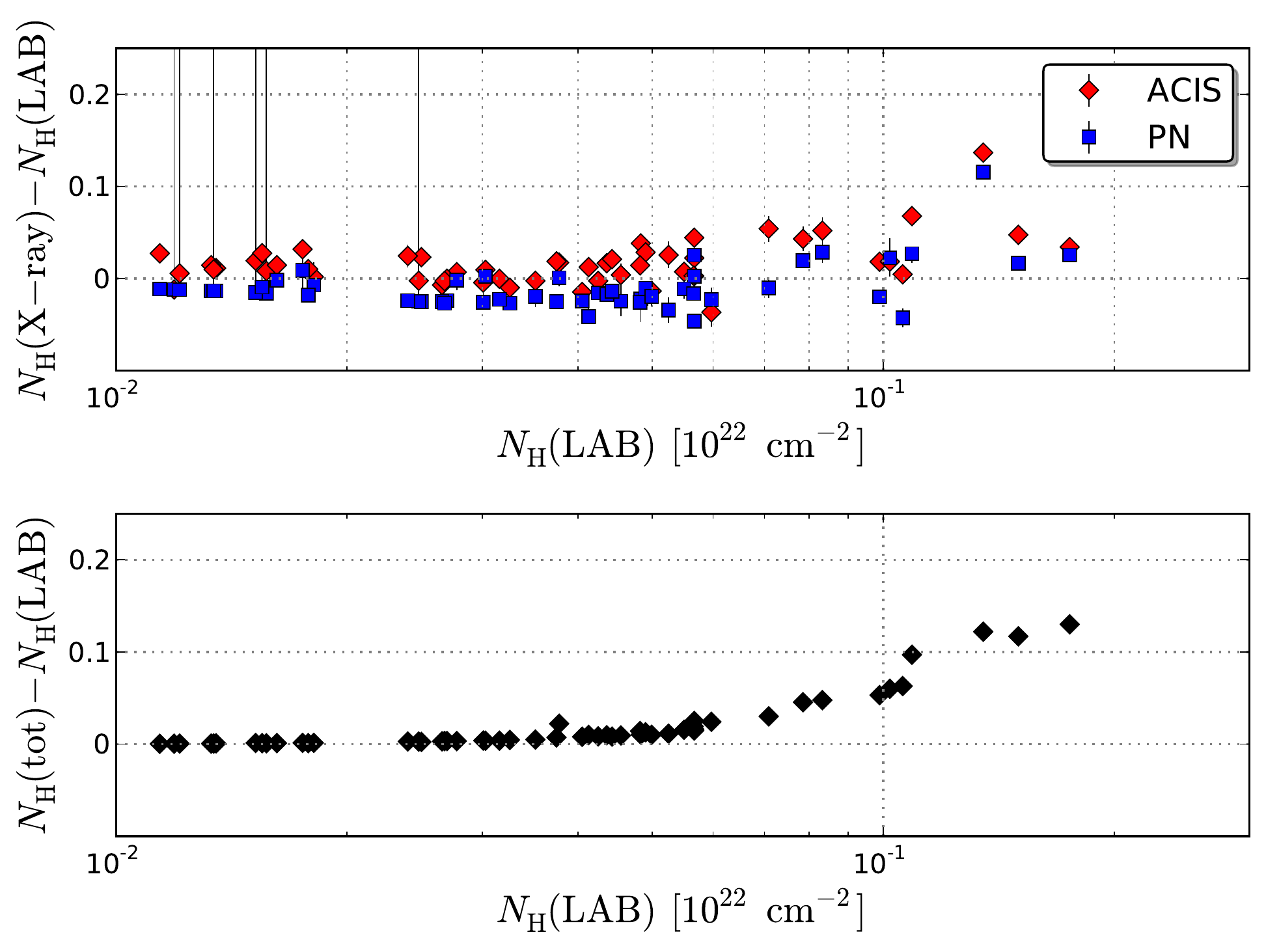}}
  \caption{\textit{Top:} Determined $N_{\rm{H}}$ values for Chandra (red diamonds) and EPIC-PN (blue squares).  \textit{Bottom:} Comparison of the $N_{\rm H, tot}$ and the LAB hydrogen column density values for the HIFLUGCS clusters. The y-axis is in units of $\SI{e22}{cm^{-2}}$ and the abundance table used is AnGr.}
  \label{fig:nhtot}
\end{figure}

We compared the obtained hydrogen column density values
with those derived including the molecular hydrogen \cite{2013MNRAS.tmp..859W} (see Fig. \ref{fig:nhtot}). The median difference and the 68\% scatter between the X-ray
derived $N_{\rm H}$ to that of \cite{2013MNRAS.tmp..859W} for ACIS and EPIC-PN
are $-0.1^{{}+1.9}_{{}-1.5}\SI{e20}{cm^{-2}}$ and
$-2.6^{{}+2.3}_{{}-1.4}\SI{e20}{cm^{-2}}$, respectively, when using
abundance table of AnGr. The fact that Chandra constrained $N_{\rm H}$ values are in better agreement with the total hydrogen column density than XMM-Newton PN, could give a hint that the Chandra calibration is more reliable, however using the Aspl abundances instead, the corresponding values are $2.0^{{}+2.8}_{{}-2.1}\SI{e20}{cm^{-2}}$ for ACIS and
$-1.4^{{}+2.8}_{{}-1.7}\SI{e20}{cm^{-2}}$ for EPIC-PN. Thus, the choice of the
abundance table (i.e. the relative abundances of the elements responsible
for absorption) and the scatter of the measurements make it difficult to interpret the results: The AnGr abundance table favors the ACIS derived column densities, while using Aspl no conclusion can be drawn.

We have demonstrated that even though the most reliable calibrated instrument can be identified theoretically by comparing the constrained hydrogen column densities with references, systematics of the abundance table that is being used, have to be taken into account.

\subsection{Cosmological Impact}
Galaxy clusters are excellent tools for cosmology, because they trace the dark matter and large scale structure in the Universe. In X-rays one can see the emission of the most massive visible component of galaxy clusters and measure this mass from the surface brightness profile (see e.g.\ \citealp{2010ApJ...722...55N}). 
Assuming hydrostatic equilibrium the total mass can be calculated. The temperature profile of the galaxy cluster enters twice in this calculation, as absolute value at a given radius and its logarithmic derivative (the former usually dominates, as shown in \citealp[Fig. 3]{2013SSRv..177..195R}). To get a good handle on the cluster masses one therefore needs accurate temperature measurements. As shown above, there exists a bias depending on the X-ray instrument used. We want to quantify what effect on the cosmological parameters the temperature difference between Chandra and XMM-Newton has. Therefore we calculated the total hydrostatic masses of all HIFLUGCS clusters by establishing for each a temperature and density profile using Chandra data. The same was done to get ``XMM-Newton Masses'' by rescaling the Chandra temperatures to ``Combined XMM-Newton'' temperatures (see Table \ref{tab:fitparam}) and with that, obtaining ``XMM-Newton Masses''. Although it is not being used for the cosmological analysis, we present here our scaling relation for the hydrostatic masses of the different instruments for comparison reasons:
\begin{equation}
\label{eq:masses}
M^{\rm XMM}_{500}  = 0.859^{{}+0.017}_{{}-0.016} \cdot \left( M^{\rm Chandra}_{500} \right)^{\num{1.00(2)}}  
\end{equation}
(see Appendix \ref{sec:relations} for more details). Since only the gradient of the density profile enters the hydrostatic equation, it is not necessary to have an accurate flux calibration for this comparison. Establishing with both, Chandra and XMM-Newton masses, a cluster mass function and comparing it to the \cite{2008ApJ...688..709T} halo mass function gives us constraints on the cosmological parameters $\Omega_{\rm M}$ and $\sigma_8$. The details about this full analysis is part of a further paper on the HIFLUGCS cluster sample and is beyond the scope of the present paper. So we use a self consistent analysis of the Chandra data, which purpose is only to quantify the relative effect of the calibration uncertainties on the cosmological parameters.

\begin{figure}
  \resizebox{\hsize}{!}{\includegraphics{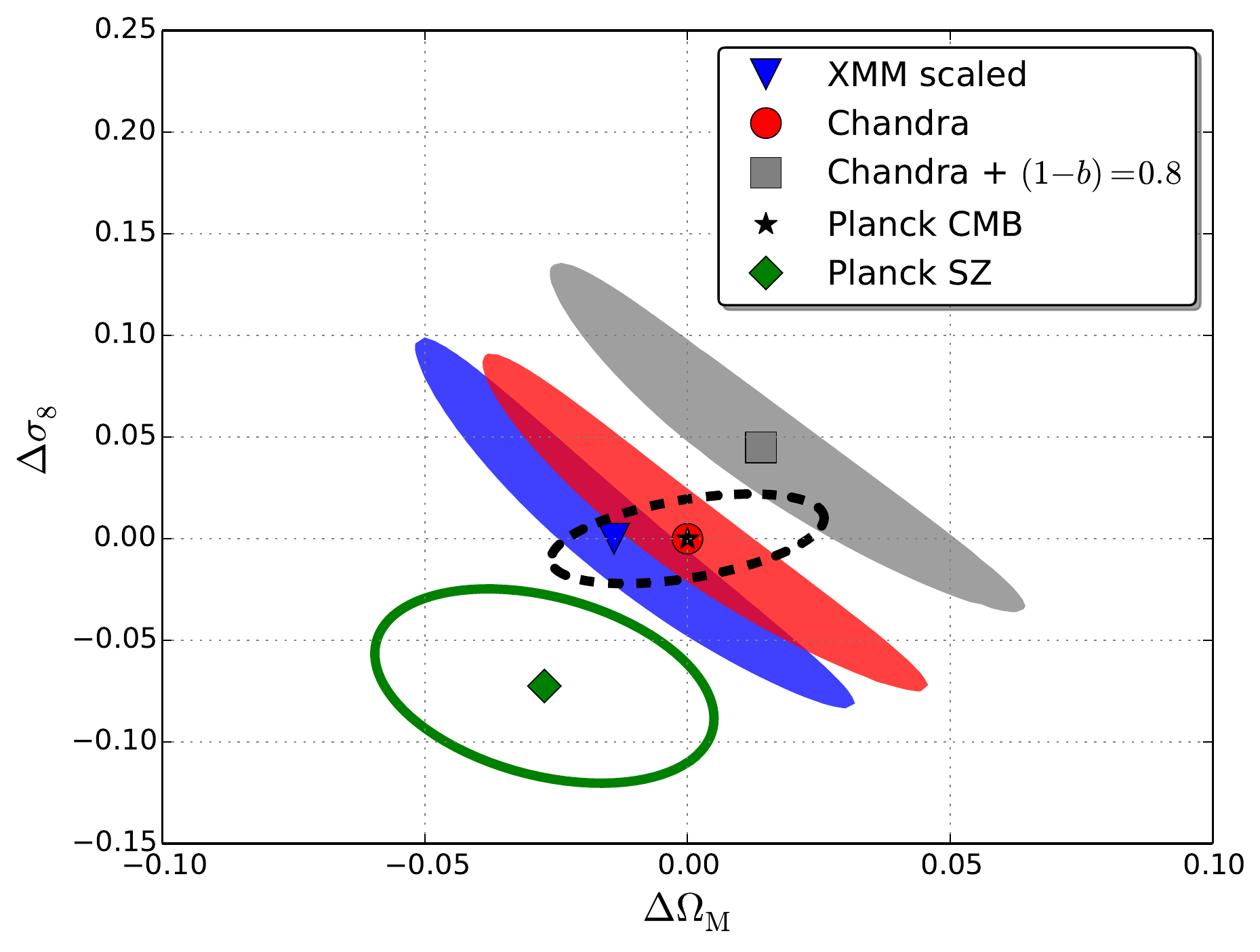}}
  \caption{Shift of the two cosmological parameters $\Omega_{\rm M}$ and $\sigma_8$ relative to the Chandra best-fit values for the X-ray analysis and relative to the Planck CMB results for the Planck analysis. The Chandra error ellipses are derived from hydrostatic masses of the HIFLUGCS clusters using Chandra temperature profiles and the Tinker halo mass function, the gray ellipse giving the results assuming a 20\% hydrostatic bias. The XMM-Newton error ellipses are derived using the Chandra derived temperature profiles and rescaling them using Eq. \ref{eq:quadfunc} and Table \ref{tab:fitparam}. The Planck error ellipses are approximated from \cite{2013arXiv1303.5080P}.}
  \label{fig:xmmscaled}
\end{figure}

The resulting shift of cosmological parameters is mainly a shift towards lower $\Omega_{\rm M}$ when using the XMM-Newton masses. We find an 
8\% lower $\Omega_{\rm M}$ and a $<1\%$ higher $\sigma_8$ for the XMM-Newton masses (see Fig. \ref{fig:xmmscaled}). Still, the 68\% confidence levels in the $\Omega_{\rm M} - \sigma_8$ plane show some overlap. For precision cosmology it is necessary to quantify all systematic biases, but it seems clear that calibration differences alone cannot account for the Planck CMB primary anisotropies and Sunyaev-Zel'dovich differences (\citealp{2013arXiv1303.5076P,2013arXiv1303.5080P}) as also shown in Figure \ref{fig:xmmscaled} where the relative difference between Planck SZ and Planck CMB are indicated by the non-filled ellipses. 
We do note that the precise shift of best-fit cosmological parameter values is expected to depend on the cluster sample under consideration.
We point out that the fact that Chandra and Planck CMB overlap in this Figure is simply due to the fact that both are arbitrarily normalized to zero for comparison reasons. This does not imply cosmological constraints from Planck CMB and Chandra galaxy clusters agree perfectly. We want to stress that the relative difference between Planck CMB and SZ best-fit values is significantly larger than the difference between Chandra and XMM-Newton derived results in the $\Omega_{\rm M}$--$\sigma_8$--plane.

In our analysis we also tested a hydrostatic bias,
\begin{equation}
(1-b) = \left< \frac{M_{\rm X-ray}}{M_{\rm true}} \right> = 0.8~,
\end{equation}
by upscaling
the Chandra masses by 25\% to simulate the strongest bias still allowed by the weak lensing (WL)  
masses determined by \cite{2014arXiv1402.3267I}.
We found a 7\% higher $\Omega_{\rm M}$ and 6\% higher $\sigma_8$ for the WL masses compared to the
Chandra analysis. This would correspond to a 14\% difference in $\Omega_{\rm M}$
and 6\% difference in $\sigma_8$ for XMM-Newton masses compared to WL. In Figure \ref{fig:xmmscaled} we indicate the contours of upscaled Chandra masses in the $\Omega_{\rm M}$ - $\sigma_8$ plane assuming $(1-b)=0.8$ by a gray square. 

In the Planck SZ contours the assumption of a uniformly varying bias in the range $[0.7, 1.0]$ enters,
while the results from \cite{2014arXiv1402.3267I} indicate the absence of any bias and allow a maximum bias of about 20\% based on the uncertainty of the intercept method. 
Without a marginalization of $(1-b)$ over the range $[0.7,1.0]$ the difference between the  cosmological parameters $\Omega_{\rm M}$ and $\sigma_8$ of the Planck primary CMB anisotropies and the SZ analyses would be even more significant, which makes it even more unlikely that the X-ray calibration uncertainties between Chandra and XMM-Newton can be responsible for this difference alone.

\section{Summary and Conclusions}\label{sec:conclusion}
We tested the calibration of Chandra and XMM-Newton using a large sample of very bright galaxy clusters. Analyzing the same regions we found significant systematic differences. First, in a direct way using the stacked residuals ratio method, we quantified the relative effective area uncertainties in the most extreme case (ACIS vs. EPIC-PN) by an increase of $\sim$20\% when moving in energy from $\SI{1}{keV}$ to $\SI{3}{keV}$. Then more indirectly but also more physically relevant, we quantified the gas temperature differences, where again the most extreme case is Chandra ACIS and XMM-Newton EPIC-PN at high cluster temperatures (i.e. EPIC-PN yielding 29\% lower temperatures at $\SI{10}{keV}$ in the full energy band.)

We showed that physical effects like multi temperature structure cannot cause the observed temperature differences.

From 
\begin{itemize}
\item  using different energy bands for spectral fitting,
\item  an energy-dependent difference of the stacked residuals ratio and
\item  hydrogen column density studies
\end{itemize}
we conclude
that systematic effective area calibration uncertainties in the soft energy band $\SIrange{0.7}{2}{keV}$ cause the observed differences. 
We provide fitting formulae to convert between Chandra and XMM-Newton using either effective area calibration files, temperatures or masses (Section \ref{sec:relations}).

To illustrate the cosmological relevance, we showed that using XMM-Newton instead of Chandra for the cluster mass function determination would result in an 8\% lower $\Omega_{\rm M}$ and $<1\%$ different $\sigma_8$. While this implies that for future high precision cluster experiments, e.g. with eROSITA (\citealp{2012MNRAS.422...44P,2012arXiv1209.3114M}), this calibration needs to be improved, it also means that this systematic uncertainty alone cannot account for the Planck CMB/SZ difference.
\begin{acknowledgements}
The authors would like to thank Douglas Applegate, Matteo Guainazzi and Fabio Gastaldello for helpful discussions. 
GS, LL, THR acknowledge support by the German Research Association (DFG) through grant RE 1462/6. THR acknowledges support by the DFG through Heisenberg grant RE 1462/5.
LL acknowledges support by the DFG through grant LO 2009/1-1 and by the German Aerospace Agency (DLR) with funds from the Ministry of Economy and Technology (BMWi) through grant 50 OR 1102. JN acknowledges a PUT 246 grant from Estonian Research Council.

\end{acknowledgements}
\bibliographystyle{aa}
\bibliography{Astro}
\begin{appendix}
\section{Tables}
\begin{table*}[htbp]
\tiny
\centering
\caption{\footnotesize List of the 63 HIFLUGCS clusters (without Abell 2244). Column (a) defines the name of the cluster as it is used in this work. Columns (b) and (c) give the coordinates in J2000 as defined in \cite{hudson_what_2009}. (d) is the redshift and (e) the hydrogen column density (both from \citealp{zhang_hiflugcs:_2010}, except Abell 478, Abell 2163 and Abell 3571, see Section \ref{sec:properties}). (f) gives the radius of the cool core region for CC clusters only (for details see Section \ref{sec:properties}). (g) and (h) give the observation IDs used and (i) and (j) are the cleaned exposure times for Chandra ACIS and XMM-Newton EPIC-PN, respectively. The * marks Chandra ACIS-S observations.\label{tab:hiflugcs}}
\begin{tabular}{cccccccccc}
\hline\hline
Cluster Name & RA& DEC & $z$ & $N_\text{H}$ & $r_{\text{core}}$ & Chandra & XMM & $t_{\rm ACIS}$ & $t_{\rm PN}$\\
& \textbf{} & \textbf{} & \textbf{} & [$10^{21}\si{cm^{-2}}$] &[$\si{kpc}$] & OBSID & OBSID & [$\si{ks}$] & [$\si{ks}$]\\
(a)&(b)&(c)&(d)&(e)&(f)&(g)&(h)&(i)&(j)\\
\hline
2A0335  & 54.6714 & 9.9672 & 0.0349 & 1.76 & 89.86 & 7939* & 0147800201 & 49.5 & 79.9 \\ 
A0085  & 10.4600 & -9.3031 & 0.0556 & 0.28 & 132.07 & 904 & 0065140101 & 38.2 & 7.4 \\ 
A0119  & 14.0668 & -1.2557 & 0.0440 & 0.33 & 0.00 & 7918 & 0505211001 & 44.8 & 6.6 \\ 
A0133  & 15.6741 & -21.8822 & 0.0569 & 0.16 & 77.87 & 9897 & 0144310101 & 68.8 & 14.2 \\ 
A0262  & 28.1926 & 36.1541 & 0.0161 & 0.64 & 27.17 & 7921* & 0109980101 & 110.5 & 12.1 \\ 
A0399  & 44.4727 & 13.0313 & 0.0715 & 1.05 & 0.00 & 3230 & 0112260101 & 48.6 & 5.1 \\ 
A0400  & 44.4233 & 6.0271 & 0.0240 & 0.83 & 0.00 & 4181 & 0404010101 & 21.5 & 14.2 \\ 
A0401  & 44.7361 & 13.5777 & 0.0748 & 0.99 & 0.00 & 14024 & 0112260301 & 134.5 & 6.5 \\ 
A0478  & 63.3548 & 10.4649 & 0.0848 & 3.00 & 122.51 & 1669* & 0109880101 & 42.2 & 40.4 \\ 
A0496  & 68.4081 & -13.2611 & 0.0328 & 0.40 & 92.02 & 4976* & 0135120201 & 58.8 & 9.4 \\ 
A0576  & 110.3761 & 55.7641 & 0.0381 & 0.55 & 0.00 & 3289* & 0205070301 & 29.0 & 6.8 \\ 
A0754  & 137.3194 & -9.6891 & 0.0528 & 0.48 & 0.00 & 10743 & 0556200501 & 93.7 & 41.0 \\ 
A1060  & 159.1781 & -27.5283 & 0.0114 & 0.50 & 0.00 & 2220 & 0206230101 & 30.6 & 24.2 \\ 
A1367  & 176.2512 & 19.6751 & 0.0216 & 0.19 & 0.00 & 514* & 0061740101 & 40.5 & 15.0 \\ 
A1644  & 194.2991 & -17.4090 & 0.0474 & 0.40 & 109.87 & 7922 & 0010420201 & 51.3 & 8.6 \\ 
A1650  & 194.6728 & -1.7619 & 0.0845 & 0.13 & 46.12 & 5823 & 0093200101 & 39.6 & 25.9 \\ 
A1651  & 194.8423 & -4.1970 & 0.0860 & 0.15 & 0.00 & 4185 & 0203020101 & 9.6 & 3.3 \\ 
A1656  & 194.8989 & 27.9597 & 0.0232 & 0.09 & 0.00 & 9714 & 0124711401 & 29.6 & 11.7 \\ 
A1736  & 201.7161 & -27.1741 & 0.0461 & 0.45 & 0.00 & 4186 & 0505210201 & 14.7 & 4.0 \\ 
A1795  & 207.2191 & 26.5925 & 0.0616 & 0.10 & 147.96 & 493* & 0097820101 & 19.6 & 22.3 \\ 
A2029  & 227.7336 & 5.7448 & 0.0767 & 0.33 & 92.48 & 4977* & 0551780401 & 77.5 & 14.4 \\ 
A2052  & 229.1813 & 7.0222 & 0.0348 & 0.27 & 40.59 & 10478* & 0109920101 & 118.3 & 17.4 \\ 
A2063  & 230.7713 & 8.6075 & 0.0354 & 0.27 & 0.00 & 6263* & 0550360101 & 16.8 & 11.1 \\ 
A2065  & 230.6222 & 27.7062 & 0.0721 & 0.31 & 103.06 & 3182 & 0112240201 & 27.7 & 7.7 \\ 
A2142  & 239.5860 & 27.2303 & 0.0899 & 0.38 & 604.90 & 5005 & 0674560201 & 44.6 & 34.0 \\ 
A2147  & 240.5699 & 15.9738 & 0.0351 & 0.28 & 0.00 & 3211 & 0505210601 & 17.9 & 5.9 \\ 
A2163  & 243.9445 & -6.1501 & 0.2010 & 2.00 & 0.00 & 1653 & 0112230601 & 71.0 & 3.8 \\ 
A2199  & 247.1597 & 39.5503 & 0.0302 & 0.09 & 32.72 & 10748 & 0008030201 & 40.4 & 10.5 \\ 
A2204  & 248.1956 & 5.5754 & 0.1523 & 0.61 & 140.25 & 7940 & 0112230301 & 76.4 & 10.9 \\ 
A2255  & 258.1423 & 64.0699 & 0.0800 & 0.23 & 0.00 & 894 & 0112260801 & 38.9 & 2.7 \\ 
A2256  & 255.8094 & 78.6500 & 0.0601 & 0.43 & 345.45 & 2419* & 0401610101 & 11.2 & 9.7 \\ 
A2589  & 350.9892 & 16.7772 & 0.0416 & 0.29 & 32.59 & 7190* & 0204180101 & 53.2 & 17.7 \\ 
A2597  & 351.3330 & -12.1243 & 0.0852 & 0.25 & 101.51 & 7329* & 0147330101 & 58.8 & 27.1 \\ 
A2634  & 354.6219 & 27.0317 & 0.0312 & 0.51 & 0.00 & 4816* & 0002960101 & 49.3 & 3.9 \\ 
A2657  & 356.2395 & 9.1919 & 0.0404 & 0.60 & 0.00 & 4941 & 0402190301 & 16.0 & 4.6 \\ 
A3112  & 49.4902 & -44.2384 & 0.0750 & 0.39 & 98.35 & 13135 & 0105660101 & 42.2 & 14.0 \\ 
A3158  & 55.7178 & -53.6321 & 0.0590 & 0.12 & 0.00 & 3712 & 0300211301 & 31.0 & 3.6 \\ 
A3266  & 67.8047 & -61.4531 & 0.0594 & 0.18 & 0.00 & 899 & 0105260901 & 29.8 & 11.9 \\ 
A3376  & 90.5360 & -39.9468 & 0.0455 & 0.44 & 0.00 & 3202 & 0151900101 & 44.3 & 16.0 \\ 
A3391  & 96.5854 & -53.6936 & 0.0531 & 0.56 & 0.00 & 4943 & 0505210401 & 16.5 & 13.4 \\ 
A3395  & 96.7073 & -54.5427 & 0.0498 & 0.74 & 0.00 & 4944 & 0400010301 & 20.3 & 17.9 \\ 
A3526  & 192.2035 & -41.3122 & 0.0103 & 0.85 & 54.41 & 4954* & 0406200101 & 87.9 & 66.4 \\ 
A3558  & 201.9870 & -31.4953 & 0.0480 & 0.40 & 112.22 & 1646* & 0107260101 & 13.8 & 28.0 \\ 
A3562  & 203.4054 & -31.6714 & 0.0499 & 0.39 & 0.00 & 4167 & 0105261301 & 19.2 & 21.2 \\ 
A3571  & 206.8680 & -32.8660 & 0.0374 & 0.42 & 0.00 & 4203* & 0086950201 & 33.6 & 18.1 \\ 
A3581  & 211.8758 & -27.0196 & 0.0214 & 0.43 & 36.62 & 12884* & 0205990101 & 84.3 & 24.8 \\ 
A3667  & 303.1778 & -56.8468 & 0.0560 & 0.46 & 26.38 & 5751 & 0206850101 & 126.8 & 39.7 \\ 
A4038  & 356.9299 & -28.1420 & 0.0283 & 0.15 & 0.00 & 4992 & 0204460101 & 33.5 & 20.0 \\ 
A4059  & 359.2539 & -34.7593 & 0.0460 & 0.12 & 42.87 & 5785* & 0109950201 & 91.7 & 12.5 \\ 
EXO0422  & 66.4635 & -8.5605 & 0.0390 & 0.81 & 47.63 & 4183 & 0300210401 & 10.0 & 24.2 \\ 
HydraA  & 139.5254 & -12.0958 & 0.0538 & 0.43 & 209.98 & 4969* & 0109980301 & 84.9 & 10.3 \\ 
IIIZw54  & 55.3235 & 15.3936 & 0.0311 & 1.47 & 0.00 & 4182 & 0505230401 & 23.5 & 20.4 \\ 
MKW3S  & 230.4656 & 7.7080 & 0.0450 & 0.29 & 106.48 & 900 & 0109930101 & 57.3 & 21.5 \\ 
MKW4  & 181.1128 & 1.8961 & 0.0200 & 0.17 & 22.19 & 3234* & 0093060101 & 29.9 & 6.7 \\ 
MKW8  & 220.1795 & 3.4660 & 0.0270 & 0.23 & 0.00 & 4942 & 0300210701 & 23.1 & 13.0 \\ 
NGC1399  & 54.6213 & -35.4502 & 0.0046 & 0.14 & 20.13 & 319* & 0400620101 & 56.0 & 51.6 \\ 
NGC1550  & 64.9082 & 2.4101 & 0.0123 & 0.98 & 20.13 & 5800* & 0152150101 & 44.3 & 15.9 \\ 
NGC4636  & 190.7080 & 2.6868 & 0.0037 & 0.18 & 16.50 & 3926 & 0111190701 & 74.2 & 42.2 \\ 
NGC5044  & 198.8495 & -16.3852 & 0.0090 & 0.51 & 51.87 & 9399* & 0037950101 & 82.7 & 10.3 \\ 
NGC507  & 20.9159 & 33.2560 & 0.0165 & 0.56 & 35.48 & 2882 & 0080540101 & 43.3 & 22.2 \\ 
RXCJ1504  & 226.0313 & -2.8047 & 0.2153 & 0.60 & 324.84 & 5793 & 0401040101 & 39.0 & 22.2 \\ 
S1101  & 348.4933 & -42.7253 & 0.0580 & 0.11 & 89.22 & 11758 & 0123900101 & 97.1 & 15.7 \\ 
ZwCl1215  & 184.4238 & 3.6551 & 0.0750 & 0.18 & 0.00 & 4184 & 0300211401 & 12.1 & 13.2 \\ 
\hline
\end{tabular}
\end{table*}
\onecolumn
{\tiny
\begin{longtable}{ccccccr}
\caption{\label{tab:temp}Best-fit temperatures and $68\%$ confidence levels for the 4 different detectors (plus EPIC combined) in the $\SIrange{0.7}{7}{keV}$ band. For cool-core clusters the annulus region was used. For details to the excluded spectra see Section \ref{sec:properties}. The last column gives the ratio of source and background count rates in the $\SIrange{0.7}{2.0}{keV}$ band.}\\
\hline
Cluster Name & $kT_{\rm ACIS}$ & $kT_{\rm MOS1}$ &$kT_{\rm MOS2}$ &$kT_{\rm PN}$&$kT_{\rm EPIC,\,Combined}$ & SC/BKG \\
& [$\si{keV}$]& [$\si{keV}$]& [$\si{keV}$]& [$\si{keV}$]& [$\si{keV}$] &\\
\hline
\endfirsthead
\caption{continued.}\\
\hline
Cluster Name & $kT_{\rm ACIS}$ & $kT_{\rm MOS1}$ &$kT_{\rm MOS2}$ &$kT_{\rm PN}$&$kT_{\rm EPIC,\,Combined}$ & SC/BKG \\
& [$\si{keV}$]& [$\si{keV}$]& [$\si{keV}$]& [$\si{keV}$]& [$\si{keV}$] &\\
\hline
\endhead
\hline
\endfoot
\parbox[0pt][1.6em][c]{0cm}{} 2A0335 & $3.99^{+0.05}_{-0.05}$ & $3.64^{+0.06}_{-0.06}$ & $3.38^{+0.04}_{-0.04}$ & $3.38^{+0.02}_{-0.02}$ & $3.45^{+0.02}_{-0.03}$ & $25.69$  \\
\parbox[0pt][1.6em][c]{0cm}{} A0085 & $6.63^{+0.12}_{-0.12}$ & $6.47^{+0.32}_{-0.32}$ & $5.75^{+0.32}_{-0.31}$ & $5.32^{+0.23}_{-0.19}$ & $5.80^{+0.17}_{-0.17}$ & $71.51$  \\
\parbox[0pt][1.6em][c]{0cm}{} A0119 & $6.45^{+0.16}_{-0.16}$ & $5.14^{+0.25}_{-0.25}$ & $5.16^{+0.28}_{-0.27}$ & $4.78^{+0.23}_{-0.23}$ & $5.02^{+0.15}_{-0.07}$ & $21.97$  \\
\parbox[0pt][1.6em][c]{0cm}{} A0133 & $4.89^{+0.09}_{-0.09}$ & $3.88^{+0.14}_{-0.14}$ & $4.02^{+0.13}_{-0.13}$ & $3.80^{+0.11}_{-0.11}$ & $3.96^{+0.05}_{-0.09}$ & $22.71$  \\
\parbox[0pt][1.6em][c]{0cm}{} A0262 & $2.38^{+0.02}_{-0.02}$ & $2.17^{+0.05}_{-0.02}$ & $2.14^{+0.04}_{-0.03}$ & $2.21^{+0.05}_{-0.05}$ & $2.19^{+0.03}_{-0.03}$ & $24.69$  \\
\parbox[0pt][1.6em][c]{0cm}{} A0399 & $7.49^{+0.16}_{-0.16}$ & $6.59^{+0.25}_{-0.25}$ & $5.90^{+0.27}_{-0.26}$ & $5.82^{+0.26}_{-0.25}$ & $6.17^{+0.14}_{-0.16}$ & $48.96$  \\
\parbox[0pt][1.6em][c]{0cm}{} A0400 & $2.60^{+0.06}_{-0.06}$ & $2.33^{+0.07}_{-0.07}$ & $2.26^{+0.07}_{-0.07}$ & $2.32^{+0.06}_{-0.06}$ & $2.31^{+0.04}_{-0.04}$ & $22.36$  \\
\parbox[0pt][1.6em][c]{0cm}{} A0401 & $10.05^{+0.10}_{-0.10}$ & $7.69^{+0.25}_{-0.25}$ & $7.25^{+0.26}_{-0.26}$ & $6.97^{+0.22}_{-0.24}$ & $7.36^{+0.13}_{-0.16}$ & $77.50$  \\
\parbox[0pt][1.6em][c]{0cm}{} A0478 & $6.95^{+0.11}_{-0.11}$ & $6.62^{+0.08}_{-0.08}$ & $6.25^{+0.07}_{-0.07}$ & $5.71^{+0.07}_{-0.07}$ & $6.14^{+0.06}_{-0.03}$ & $49.05$  \\
\parbox[0pt][1.6em][c]{0cm}{} A0496 & $5.18^{+0.07}_{-0.07}$ & $4.44^{+0.18}_{-0.15}$ & $4.41^{+0.17}_{-0.15}$ & $4.23^{+0.13}_{-0.11}$ & $4.39^{+0.11}_{-0.08}$ & $41.73$  \\
\parbox[0pt][1.6em][c]{0cm}{} A0576 & $4.42^{+0.08}_{-0.08}$ & $3.72^{+0.16}_{-0.16}$ & $3.47^{+0.15}_{-0.15}$ & $3.81^{+0.13}_{-0.14}$ & $3.69^{+0.09}_{-0.07}$ & $29.60$  \\
\parbox[0pt][1.6em][c]{0cm}{} A0754 & $9.72^{+0.21}_{-0.21}$ & $7.45^{+0.24}_{-0.24}$ & $8.22^{+0.23}_{-0.23}$ & $6.36^{+0.14}_{-0.14}$ & $7.24^{+0.13}_{-0.11}$ & $75.28$  \\
\parbox[0pt][1.6em][c]{0cm}{} A1060 & $3.69^{+0.05}_{-0.05}$ & $3.33^{+0.04}_{-0.04}$ & $3.19^{+0.03}_{-0.03}$ & $3.06^{+0.03}_{-0.03}$ & $3.17^{+0.03}_{-0.01}$ & $92.29$  \\
\parbox[0pt][1.6em][c]{0cm}{} A1367 & $3.54^{+0.10}_{-0.10}$ & $2.52^{+0.15}_{-0.15}$ & $2.71^{+0.15}_{-0.14}$ & $2.41^{+0.14}_{-0.15}$ & $2.56^{+0.08}_{-0.08}$ & $13.92$  \\
\parbox[0pt][1.6em][c]{0cm}{} A1644 & $5.31^{+0.14}_{-0.13}$ & $4.92^{+0.34}_{-0.34}$ & $4.20^{+0.28}_{-0.25}$ & $4.63^{+0.28}_{-0.29}$ & $4.61^{+0.19}_{-0.17}$ & $22.80$  \\
\parbox[0pt][1.6em][c]{0cm}{} A1650 & $6.43^{+0.10}_{-0.10}$ & $5.41^{+0.13}_{-0.10}$ & $5.09^{+0.10}_{-0.10}$ & $4.88^{+0.08}_{-0.08}$ & $5.14^{+0.05}_{-0.05}$ & $55.96$  \\
\parbox[0pt][1.6em][c]{0cm}{} A1651 & $7.07^{+0.25}_{-0.25}$ & $6.32^{+0.20}_{-0.20}$ & $5.88^{+0.24}_{-0.24}$ & $5.92^{+0.20}_{-0.21}$ & $6.09^{+0.12}_{-0.12}$ & $97.23$  \\
\parbox[0pt][1.6em][c]{0cm}{} A1656 & $9.68^{+0.21}_{-0.21}$ & $8.40^{+0.21}_{-0.21}$ & $7.39^{+0.21}_{-0.21}$ & $6.62^{+0.11}_{-0.11}$ & $7.44^{+0.09}_{-0.13}$ & $88.21$  \\
\parbox[0pt][1.6em][c]{0cm}{} A1736 & $3.67^{+0.13}_{-0.13}$ & $3.57^{+0.18}_{-0.18}$ & $3.07^{+0.15}_{-0.15}$ & $3.10^{+0.17}_{-0.17}$ & $3.26^{+0.08}_{-0.09}$ & $28.06$  \\
\parbox[0pt][1.6em][c]{0cm}{} A1795 & $6.06^{+0.16}_{-0.16}$ & $6.18^{+0.14}_{-0.14}$ & $5.56^{+0.14}_{-0.14}$ & $5.32^{+0.09}_{-0.09}$ & $5.70^{+0.08}_{-0.08}$ & $55.67$  \\
\parbox[0pt][1.6em][c]{0cm}{} A2029 & $8.75^{+0.11}_{-0.11}$ & $7.74^{+0.16}_{-0.16}$ & $7.09^{+0.15}_{-0.15}$ & $6.36^{+0.11}_{-0.11}$ & $7.03^{+0.07}_{-0.10}$ & $71.64$  \\
\parbox[0pt][1.6em][c]{0cm}{} A2052 & $3.51^{+0.03}_{-0.03}$ & $3.16^{+0.05}_{-0.05}$ & $3.00^{+0.05}_{-0.05}$ & $2.96^{+0.04}_{-0.04}$ & $3.04^{+0.03}_{-0.03}$ & $31.28$  \\
\parbox[0pt][1.6em][c]{0cm}{} A2063 & $3.87^{+0.06}_{-0.06}$ & $3.75^{+0.07}_{-0.07}$ & $3.58^{+0.06}_{-0.06}$ & $3.31^{+0.05}_{-0.05}$ & $3.53^{+0.04}_{-0.03}$ & $44.95$  \\
\parbox[0pt][1.6em][c]{0cm}{} A2065 & $6.68^{+0.17}_{-0.17}$ & $5.49^{+0.31}_{-0.24}$ & $4.61^{+0.23}_{-0.23}$ & $4.24^{+0.18}_{-0.16}$ & $4.81^{+0.13}_{-0.15}$ & $36.67$  \\
\parbox[0pt][1.6em][c]{0cm}{} A2142 &  - &  - &  - &  - &  - &  -  \\
\parbox[0pt][1.6em][c]{0cm}{} A2147 & $4.50^{+0.15}_{-0.15}$ & $4.15^{+0.16}_{-0.16}$ & $4.32^{+0.23}_{-0.17}$ & $4.09^{+0.15}_{-0.15}$ & $4.19^{+0.07}_{-0.09}$ & $39.33$  \\
\parbox[0pt][1.6em][c]{0cm}{} A2163 & $11.73^{+0.25}_{-0.25}$ & $9.73^{+0.45}_{-0.45}$ & $9.92^{+0.45}_{-0.45}$ & $8.06^{+0.35}_{-0.35}$ & $9.40^{+0.27}_{-0.29}$ & $65.74$  \\
\parbox[0pt][1.6em][c]{0cm}{} A2199 & $4.77^{+0.04}_{-0.04}$ & $4.31^{+0.08}_{-0.06}$ & $4.02^{+0.07}_{-0.07}$ & $3.95^{+0.05}_{-0.05}$ & $4.10^{+0.03}_{-0.04}$ & $103.19$  \\
\parbox[0pt][1.6em][c]{0cm}{} A2204 & $9.97^{+0.24}_{-0.24}$ & $7.58^{+0.34}_{-0.33}$ & $7.03^{+0.32}_{-0.32}$ & $6.80^{+0.35}_{-0.20}$ & $7.22^{+0.17}_{-0.18}$ & $28.37$  \\
\parbox[0pt][1.6em][c]{0cm}{} A2255 & $7.22^{+0.23}_{-0.23}$ & $6.88^{+0.57}_{-0.40}$ & $6.15^{+0.49}_{-0.48}$ & $6.54^{+0.56}_{-0.48}$ & $6.62^{+0.25}_{-0.25}$ & $25.73$  \\
\parbox[0pt][1.6em][c]{0cm}{} A2256 &  - &  - &  - &  - &  - &  -  \\
\parbox[0pt][1.6em][c]{0cm}{} A2589 & $3.91^{+0.04}_{-0.04}$ & $3.51^{+0.08}_{-0.08}$ & $3.34^{+0.05}_{-0.06}$ & $3.38^{+0.05}_{-0.05}$ & $3.40^{+0.03}_{-0.02}$ & $25.80$  \\
\parbox[0pt][1.6em][c]{0cm}{} A2597 & $4.08^{+0.07}_{-0.07}$ & $3.50^{+0.08}_{-0.08}$ & $3.36^{+0.06}_{-0.06}$ & $3.04^{+0.06}_{-0.06}$ & $3.30^{+0.04}_{-0.03}$ & $14.65$  \\
\parbox[0pt][1.6em][c]{0cm}{} A2634 & $3.91^{+0.10}_{-0.10}$ & $3.85^{+0.30}_{-0.30}$ & $3.24^{+0.22}_{-0.21}$ & $2.63^{+0.25}_{-0.16}$ & $3.22^{+0.13}_{-0.08}$ & $10.60$  \\
\parbox[0pt][1.6em][c]{0cm}{} A2657 & $4.01^{+0.08}_{-0.08}$ & $3.97^{+0.10}_{-0.10}$ & $3.52^{+0.10}_{-0.09}$ & $3.96^{+0.12}_{-0.12}$ & $3.81^{+0.06}_{-0.06}$ & $45.92$  \\
\parbox[0pt][1.6em][c]{0cm}{} A3112 & $5.45^{+0.12}_{-0.09}$ & $4.42^{+0.14}_{-0.15}$ & $3.97^{+0.09}_{-0.10}$ & $3.79^{+0.08}_{-0.08}$ & $4.00^{+0.06}_{-0.04}$ & $33.10$  \\
\parbox[0pt][1.6em][c]{0cm}{} A3158 & $6.01^{+0.10}_{-0.10}$ & $5.54^{+0.23}_{-0.18}$ & $4.92^{+0.15}_{-0.15}$ & $4.84^{+0.15}_{-0.15}$ & $5.11^{+0.10}_{-0.08}$ & $83.88$  \\
\parbox[0pt][1.6em][c]{0cm}{} A3266 & $9.99^{+0.26}_{-0.26}$ & $7.80^{+0.29}_{-0.29}$ & $8.15^{+0.29}_{-0.29}$ & $6.81^{+0.27}_{-0.18}$ & $7.62^{+0.17}_{-0.17}$ & $93.71$  \\
\parbox[0pt][1.6em][c]{0cm}{} A3376 & $4.78^{+0.14}_{-0.14}$ & $3.68^{+0.18}_{-0.18}$ & $3.52^{+0.19}_{-0.16}$ & $3.25^{+0.11}_{-0.11}$ & $3.43^{+0.11}_{-0.07}$ & $17.92$  \\
\parbox[0pt][1.6em][c]{0cm}{} A3391 & $6.62^{+0.22}_{-0.22}$ & $5.86^{+0.21}_{-0.21}$ & $5.61^{+0.20}_{-0.19}$ & $5.22^{+0.13}_{-0.13}$ & $5.54^{+0.13}_{-0.09}$ & $34.57$  \\
\parbox[0pt][1.6em][c]{0cm}{} A3395 & $5.27^{+0.24}_{-0.22}$ & $4.73^{+0.25}_{-0.25}$ & $4.91^{+0.25}_{-0.25}$ & $4.50^{+0.20}_{-0.20}$ & $4.72^{+0.12}_{-0.14}$ & $15.67$  \\
\parbox[0pt][1.6em][c]{0cm}{} A3526 &  - &  - &  - &  - &  - &  -  \\
\parbox[0pt][1.6em][c]{0cm}{} A3558 & $7.42^{+0.27}_{-0.28}$ & $5.68^{+0.15}_{-0.15}$ & $5.37^{+0.13}_{-0.11}$ & $5.36^{+0.10}_{-0.09}$ & $5.51^{+0.08}_{-0.08}$ & $55.17$  \\
\parbox[0pt][1.6em][c]{0cm}{} A3562 & $4.97^{+0.11}_{-0.11}$ & $4.29^{+0.08}_{-0.07}$ & $4.13^{+0.08}_{-0.08}$ & $4.03^{+0.07}_{-0.07}$ & $4.16^{+0.04}_{-0.04}$ & $50.44$  \\
\parbox[0pt][1.6em][c]{0cm}{} A3571 & $8.10^{+0.08}_{-0.08}$ & $6.69^{+0.08}_{-0.08}$ & $6.37^{+0.08}_{-0.08}$ & $6.09^{+0.06}_{-0.06}$ & $6.36^{+0.06}_{-0.03}$ & $135.09$  \\
\parbox[0pt][1.6em][c]{0cm}{} A3581 & $2.06^{+0.01}_{-0.01}$ & $1.86^{+0.03}_{-0.03}$ & $1.84^{+0.03}_{-0.03}$ & $1.76^{+0.03}_{-0.03}$ & $1.83^{+0.01}_{-0.02}$ & $17.71$  \\
\parbox[0pt][1.6em][c]{0cm}{} A3667 & $6.65^{+0.05}_{-0.05}$ & $5.48^{+0.13}_{-0.11}$ & $5.37^{+0.12}_{-0.10}$ & $5.02^{+0.07}_{-0.07}$ & $5.25^{+0.05}_{-0.05}$ & $58.63$  \\
\parbox[0pt][1.6em][c]{0cm}{} A4038 & $3.38^{+0.03}_{-0.03}$ & $3.09^{+0.04}_{-0.04}$ & $3.00^{+0.03}_{-0.03}$ & $2.90^{+0.03}_{-0.03}$ & $2.99^{+0.02}_{-0.02}$ & $109.43$  \\
\parbox[0pt][1.6em][c]{0cm}{} A4059 & $4.61^{+0.04}_{-0.04}$ & $3.93^{+0.08}_{-0.08}$ & $3.96^{+0.08}_{-0.08}$ & $3.89^{+0.07}_{-0.07}$ & $3.96^{+0.04}_{-0.04}$ & $40.47$  \\
\parbox[0pt][1.6em][c]{0cm}{} EXO0422 & $3.43^{+0.12}_{-0.09}$ & $3.24^{+0.05}_{-0.05}$ & $3.19^{+0.05}_{-0.05}$ & $3.03^{+0.04}_{-0.04}$ & $3.15^{+0.03}_{-0.02}$ & $48.04$  \\
\parbox[0pt][1.6em][c]{0cm}{} HydraA &  - &  - &  - &  - &  - &  -  \\
\parbox[0pt][1.6em][c]{0cm}{} IIIZw54 & $2.80^{+0.06}_{-0.06}$ & $2.70^{+0.05}_{-0.05}$ & $2.57^{+0.05}_{-0.05}$ & $2.60^{+0.04}_{-0.04}$ & $2.63^{+0.02}_{-0.03}$ & $34.60$  \\
\parbox[0pt][1.6em][c]{0cm}{} MKW3S & $4.06^{+0.06}_{-0.06}$ & $3.48^{+0.10}_{-0.10}$ & $3.31^{+0.07}_{-0.07}$ & $3.22^{+0.06}_{-0.06}$ & $3.32^{+0.05}_{-0.03}$ & $57.74$  \\
\parbox[0pt][1.6em][c]{0cm}{} MKW4 & $2.14^{+0.02}_{-0.02}$ & $1.94^{+0.06}_{-0.06}$ & $1.95^{+0.05}_{-0.06}$ & $1.86^{+0.06}_{-0.06}$ & $1.93^{+0.03}_{-0.02}$ & $17.30$  \\
\parbox[0pt][1.6em][c]{0cm}{} MKW8 & $3.51^{+0.12}_{-0.11}$ & $3.16^{+0.11}_{-0.11}$ & $2.66^{+0.09}_{-0.08}$ & $2.90^{+0.09}_{-0.09}$ & $2.91^{+0.06}_{-0.04}$ & $18.74$  \\
\parbox[0pt][1.6em][c]{0cm}{} NGC1399 &  - &  - &  - &  - &  - &  -  \\
\parbox[0pt][1.6em][c]{0cm}{} NGC1550 & $1.55^{+0.02}_{-0.02}$ & $1.54^{+0.04}_{-0.04}$ & $1.48^{+0.04}_{-0.05}$ & $1.50^{+0.03}_{-0.03}$ & $1.52^{+0.02}_{-0.02}$ & $28.38$  \\
\parbox[0pt][1.6em][c]{0cm}{} NGC4636 &  - &  - &  - &  - &  - &  -  \\
\parbox[0pt][1.6em][c]{0cm}{} NGC5044 &  - &  - &  - &  - &  - &  -  \\
\parbox[0pt][1.6em][c]{0cm}{} NGC507 & $1.54^{+0.02}_{-0.02}$ & $1.51^{+0.03}_{-0.03}$ & $1.44^{+0.04}_{-0.04}$ & $1.53^{+0.02}_{-0.02}$ & $1.51^{+0.01}_{-0.02}$ & $21.37$  \\
\parbox[0pt][1.6em][c]{0cm}{} RXCJ1504 & $9.81^{+0.80}_{-0.79}$ & $7.46^{+0.54}_{-0.54}$ & $5.97^{+0.38}_{-0.37}$ & $5.99^{+0.31}_{-0.31}$ & $6.40^{+0.20}_{-0.16}$ & $8.87$  \\
\parbox[0pt][1.6em][c]{0cm}{} S1101 & $2.88^{+0.04}_{-0.04}$ & $2.54^{+0.06}_{-0.06}$ & $2.30^{+0.07}_{-0.07}$ & $2.37^{+0.06}_{-0.06}$ & $2.42^{+0.04}_{-0.03}$ & $20.09$  \\
\parbox[0pt][1.6em][c]{0cm}{} ZwCl1215 & $7.17^{+0.30}_{-0.30}$ & $6.57^{+0.16}_{-0.16}$ & $6.07^{+0.17}_{-0.17}$ & $5.60^{+0.15}_{-0.15}$ & $6.09^{+0.09}_{-0.09}$ & $59.56$  \\
\end{longtable}
}
\begin{table}[h]
  \centering
  \footnotesize
  \caption{Spline parameters (e.g. \citealp[Chapter 3.3]{1992nrca.book.....P}) for the stacked residuals ratios. $y$ denotes the stacked residuals ratio value at a given energy and $y''$ the second derivative. All splines are normalized to unity at $\SI{1.1}{keV}$.}  
    \begin{tabular}[tbp]{c|cr|cr|cr}
    \parbox[0pt][1.6em][c]{0cm}{} Energy & \multicolumn{2}{c}{ACIS/PN} & \multicolumn{2}{c}{MOS1/PN} & \multicolumn{2}{c}{MOS2/PN} \\
     \parbox[0pt][1.6em][c]{0cm}{} [keV] & $y$ & $y''$ & $y$ & $y''$ & $y$ & $y''$ \\
    \hline
    0.54 & 1.01 & -14.69 & 0.99 & -9.32 & 1.00 & -11.37\\
    0.62 & 1.04 & -0.81 & 1.06 & -1.23 & 1.05 & -1.08\\
    0.71 & 0.99 & 0.02 & 1.03 & 0.16 & 1.02 & 0.32\\
    0.81 & 0.96 & 0.49 & 1.02 & 0.20 & 1.02 & -0.02\\
    0.94 & 0.98 & -0.06 & 1.02 & -0.34 & 1.01 & -0.06\\
    1.08 & 1.00 & 0.02 & 1.00 & 0.56 & 1.00 & 0.35\\
    1.24 & 1.02 & 0.07 & 1.05 & -0.23 & 1.04 & -0.16\\
    1.42 & 1.06 & -0.10 & 1.07 & 0.01 & 1.06 & 0.02\\
    1.63 & 1.07 & -0.08 & 1.08 & -0.18 & 1.08 & -0.23\\
    1.88 & 1.08 & 0.09 & 1.05 & 0.14 & 1.05 & 0.13\\
    2.15 & 1.11 & 0.03 & 1.07 & 0.03 & 1.05 & -0.00\\
    2.48 & 1.15 & -0.06 & 1.09 & -0.07 & 1.05 & -0.00\\
    2.84 & 1.17 & -0.09 & 1.08 & 0.02 & 1.04 & -0.07\\
    3.27 & 1.15 & 0.03 & 1.09 & -0.00 & 1.02 & 0.12\\
    3.76 & 1.15 & -0.01 & 1.10 & 0.03 & 1.05 & 0.02\\
    4.32 & 1.15 & 0.09 & 1.12 & -0.06 & 1.08 & -0.12\\
    4.96 & 1.18 & -0.17 & 1.10 & 0.01 & 1.04 & 0.05\\
    5.70 & 1.13 & 0.15 & 1.09 & -0.01 & 1.01 & -0.12\\
    6.55 & 1.16 & -0.17 & 1.08 & 0.05 & 0.95 & 0.32\\
  \end{tabular}
  \label{tab:spline}
\end{table}
\section{Temperature Comparison}\label{sec:app_temp}
\begin{figure}[h]
  \resizebox{0.90\hsize}{!}{\includegraphics{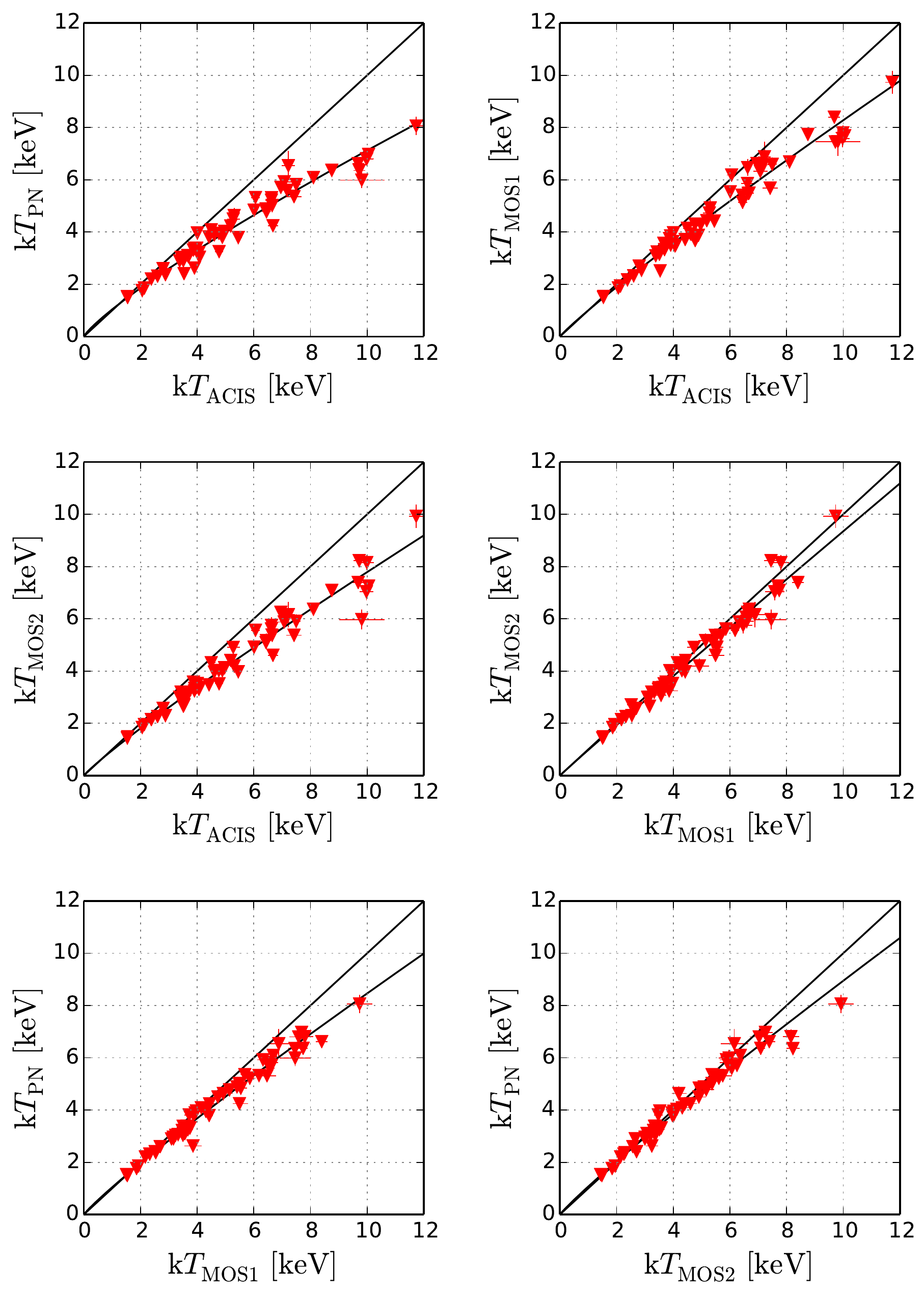}}
  \caption{Best-fit temperatures of the HIFLUGCS clusters in an isothermal region for all detector combinations in the $\SIrange{0.7}{7.0}{keV}$ energy band and with $N_{\rm H}$ frozen to the radio value of the LAB survey. The parameters of the best-fit powerlaw (black line) are also shown in Fig.~\ref{fig:deg_log}.}
  \label{fig:t_all_log_full}
\end{figure}
\begin{figure}[h]
  \resizebox{0.90\hsize}{!}{\includegraphics{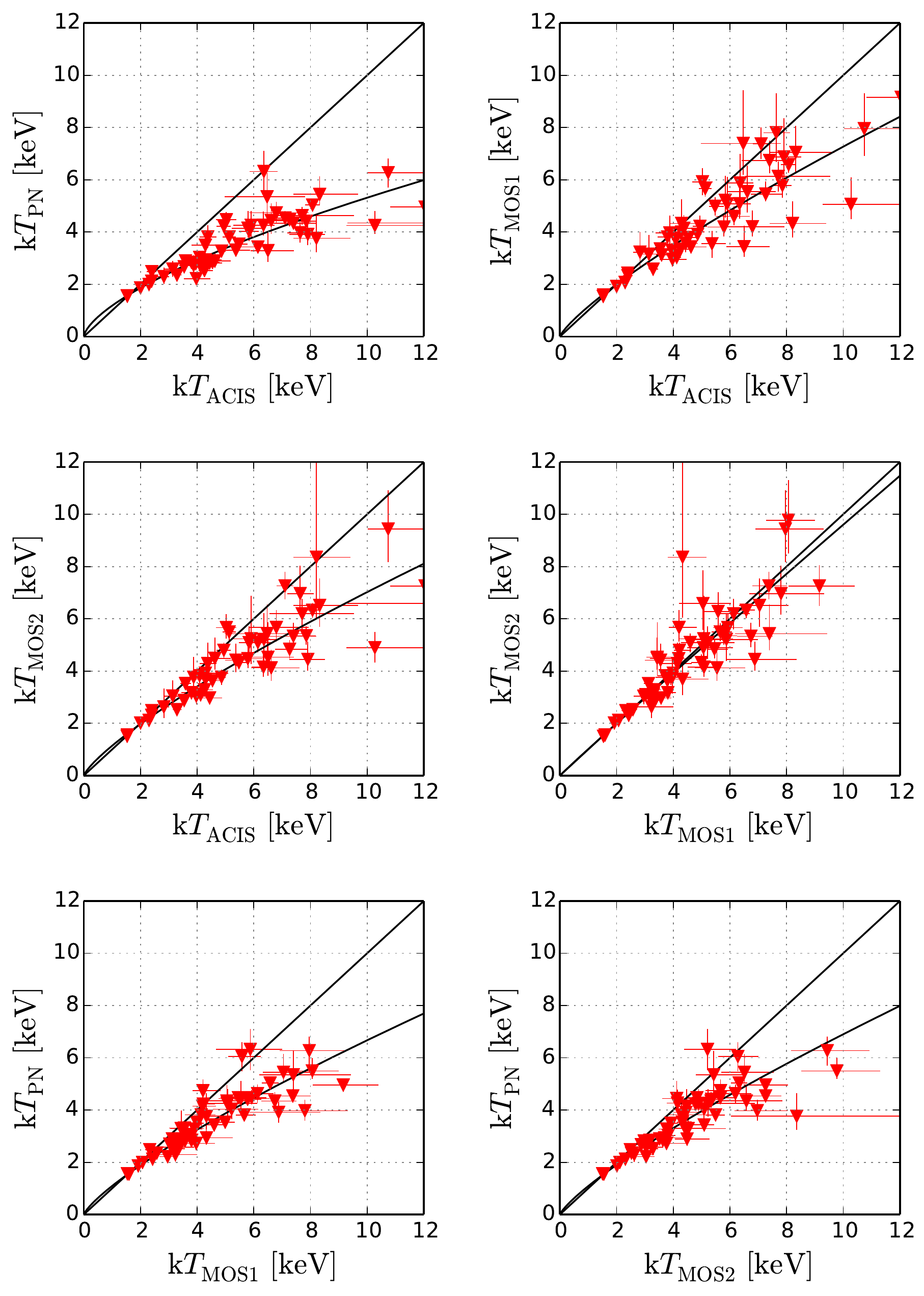}}
  \caption{Same as Figure \ref{fig:t_all_log_full} but for the $\SIrange{0.7}{2.0}{keV}$ band.}
  \label{fig:t_all_log_soft}
\end{figure}
\begin{figure}[h]
  \resizebox{0.90\hsize}{!}{\includegraphics{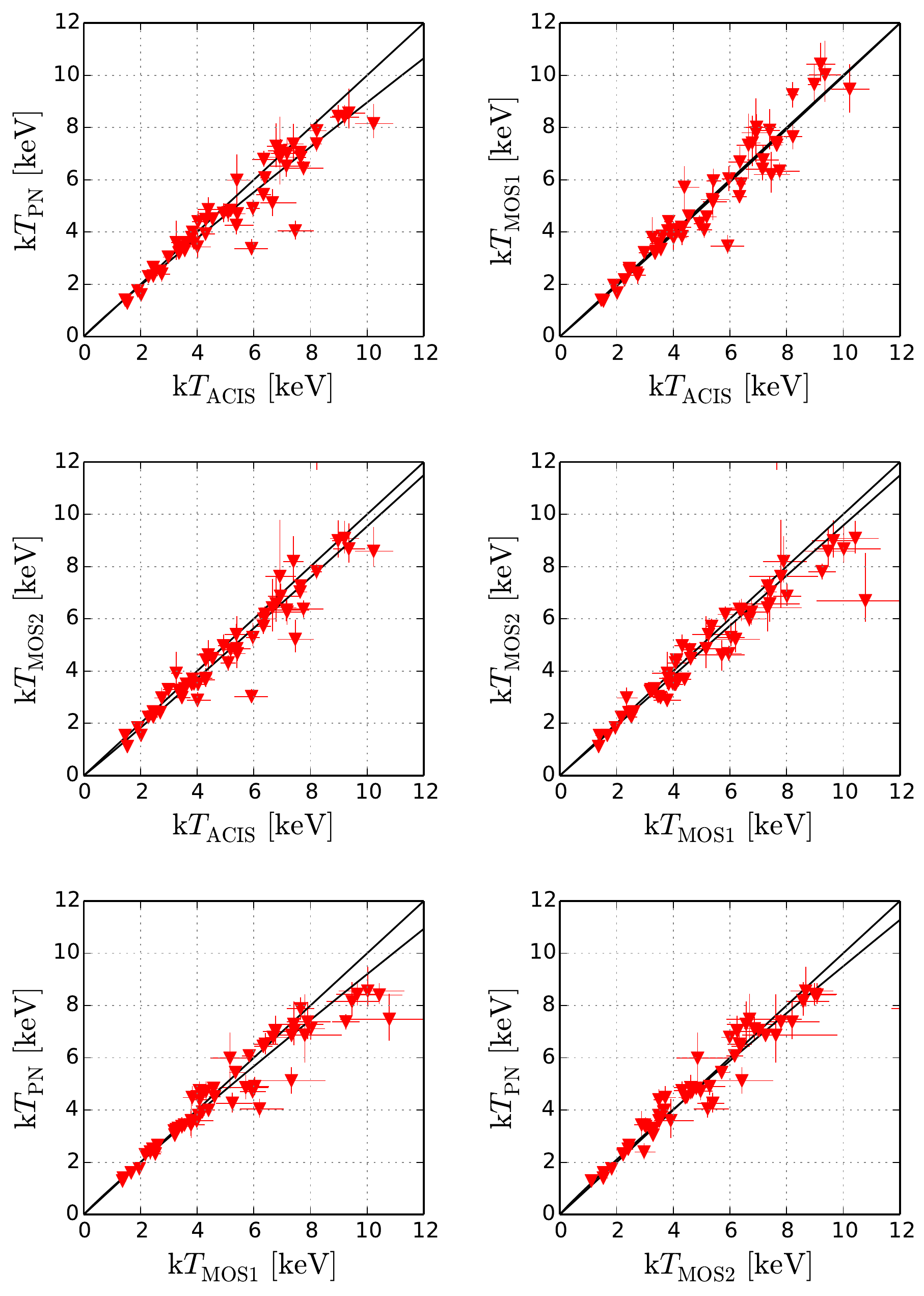}}
  \caption{Same as Figure \ref{fig:t_all_log_full} but for the $\SIrange{2.0}{7.0}{keV}$ band.}
  \label{fig:t_all_log_hard}
\end{figure}
\begin{figure}[h]
  \resizebox{0.90\hsize}{!}{\includegraphics{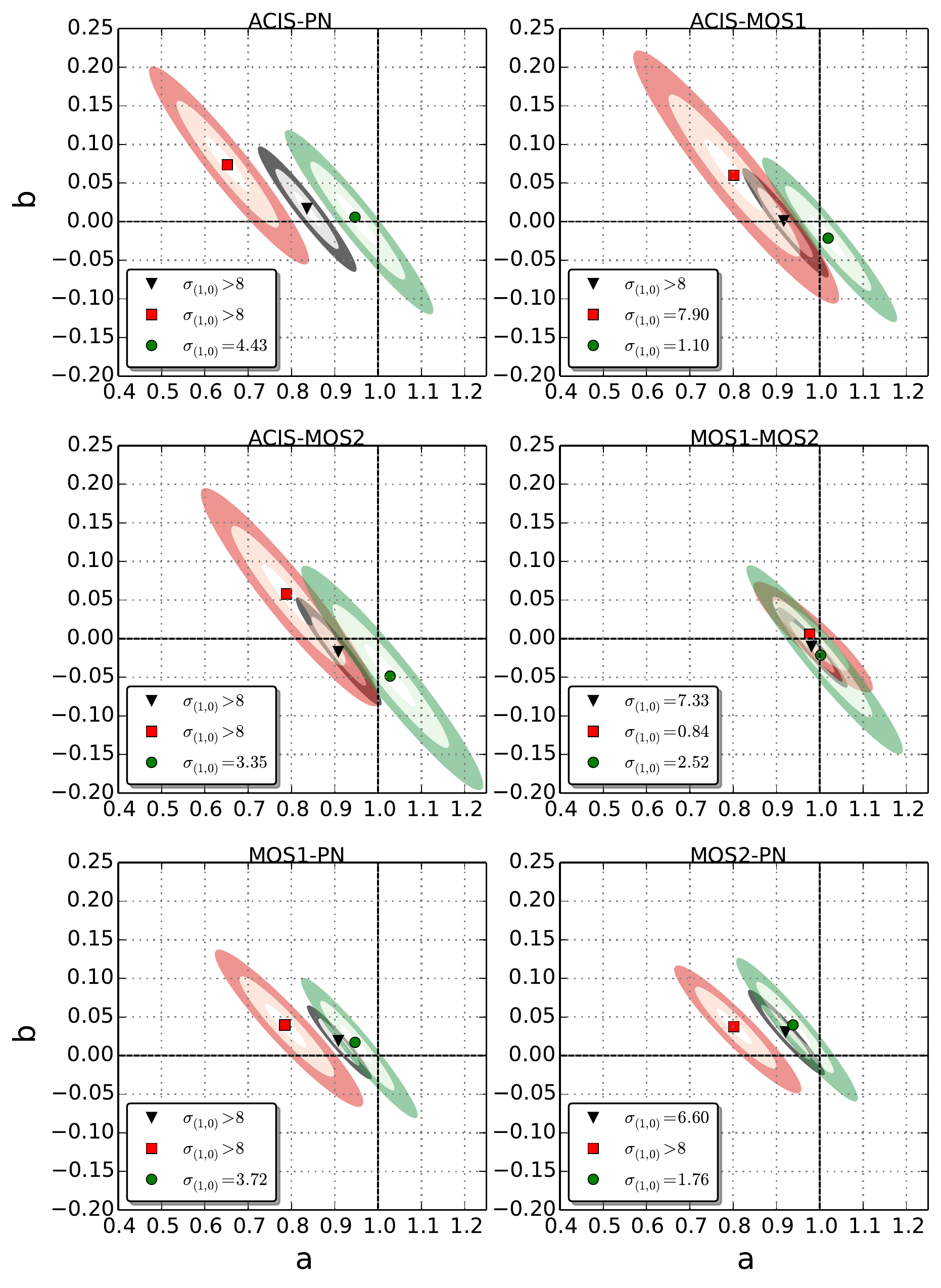}}
  \caption{Fit parameter (see Eq. \ref{eq:quadfunc}) degeneracy for the 1-, 3- and 5-$\sigma$ levels of the different detector combinations for the full (gray triangle), soft (red square) and hard (green circle) energy band. Equality of temperatures for two instruments is given for $a=1$ and $b=0$. The deviation in $\sigma$ from equality is given in the legend.}
  \label{fig:deg_log}
\end{figure}
\begin{figure}[h]
\centering
  \resizebox{0.90\hsize}{!}{\includegraphics{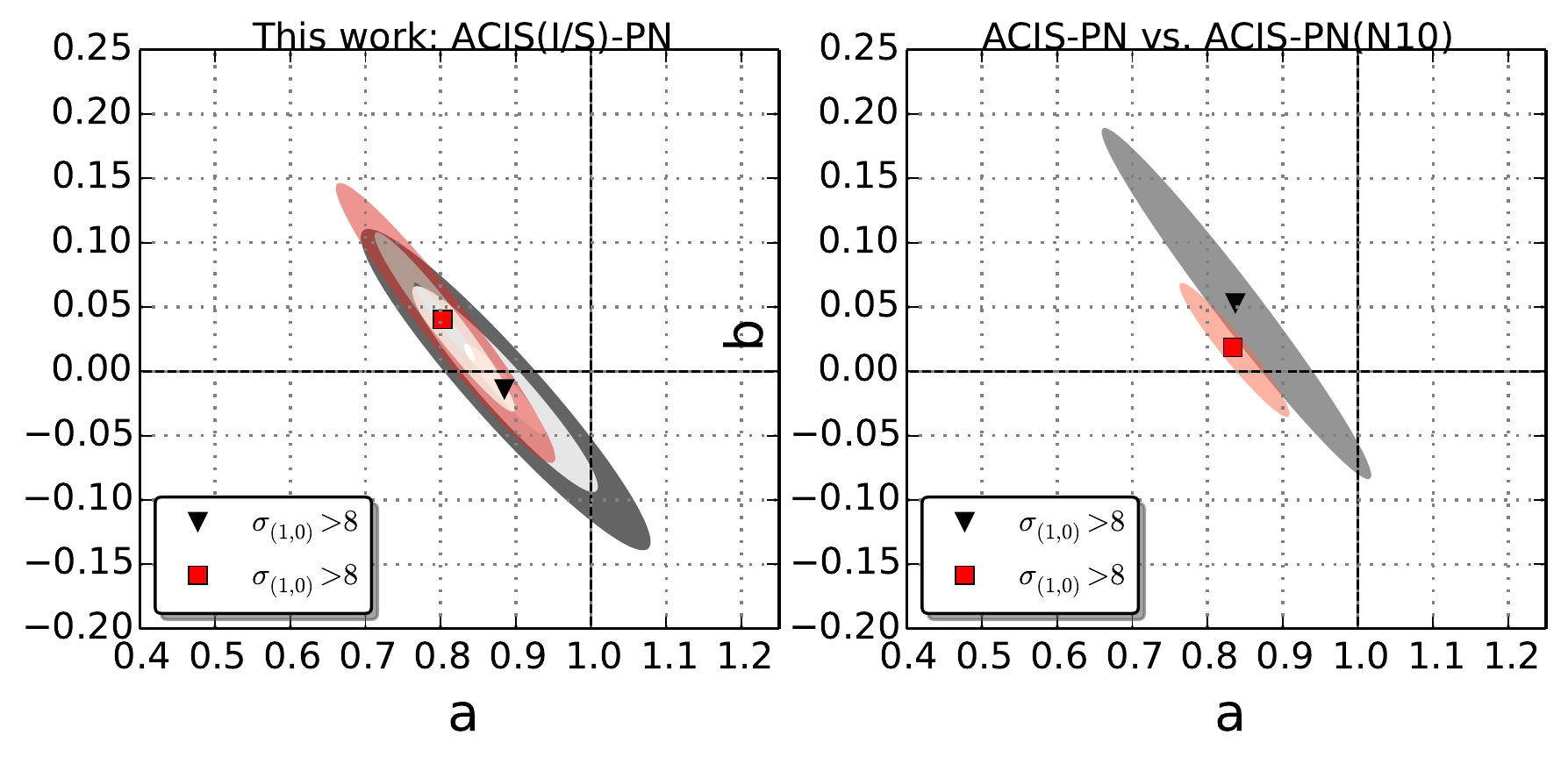}}
  \caption{\textit{Left: }Fit parameter degeneracy for the 1-, 3- and 5-$\sigma$ levels of ACIS - PN for the $\SIrange{0.7}{7}{keV}$ energy band. ACIS-I (red square) and ACIS-S (black triangle) HIFLUGCS subsamples are shown versus the PN data (for all ACIS data combined see Fig. \ref{fig:deg_log}, top left panel, gray ellipse). Equality of temperatures for two instruments is given for $a=1$ and $b=0$. The deviation in $\sigma$ from equality is given in the legend. \textit{Right: } As in the left panel but only 3-$\sigma$ levels (gray and red ellipses) of ACIS - PN with temperatures taken from N10 ($0.5-7\,\si{keV}$ energy band, black triangle) and complete ACIS - PN of this work ($0.7 - 7\,\si{keV}$ energy band, red square). }
  \label{fig:deg2_log}
\end{figure}
\clearpage
\section{Self consistency test}\label{ch:self-consistency}
In Section \ref{ch:temps} we have compared the best-fit temperatures of different instruments in the same energy band. In the case of purely isothermal emission and a very accurately calibrated effective area, the temperatures obtained in different energy bands of the same instrument should also be equal. So by comparing temperatures of one instrument in different bands it is possible to quantify the absolute calibration uncertainties, if the assumption of isothermal emission is fulfilled. 

\begin{figure}[!h]
\centering
  \resizebox{0.90\hsize}{!}{\includegraphics{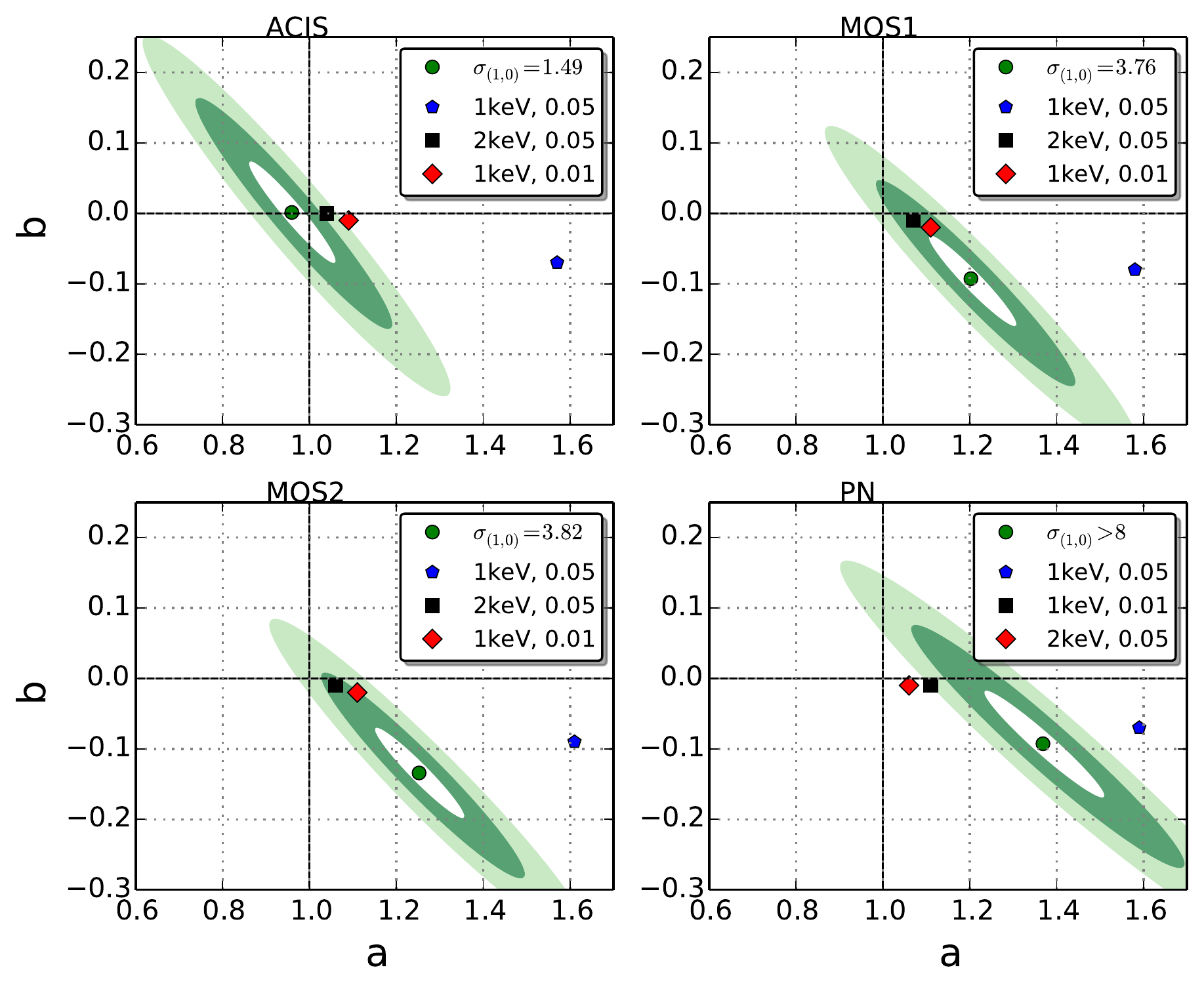}}
  \caption{Fit parameter degeneracy for the 1-,3- and 5-$\sigma$ levels of each instrument in different energy bands. The green contours (circles) refer to the soft vs hard band. Equality of temperatures for two bands is given for $a=1$ and $b=0$. The deviation in $\sigma$ from equality is given in the legend as well as the expectations (blue pentagon, black square and red diamond) for a multi temperature ICM with the parameters (temperature, emission measure ratio) of the cold component given.}
  \label{fig:deg3_log}
\end{figure}

In Figure \ref{fig:deg3_log} we demonstrate how self consistent the instruments are in terms of soft and hard band temperatures. We also quantify the expected deviation from equality of soft and hard band temperatures for a given two temperature plasma.

Only for ACIS the temperature deviation (quoted $\sigma$ value in the legend of Fig. \ref{fig:deg3_log}) is less than $3\sigma$ in comparing the soft and hard energy bands (ACIS has lowest $\sigma$ and is closest to the (1,0) equality point in Fig. \ref{fig:deg3_log}). 
By performing simulations similar to those shown in \ref{ch:icm_simu} we can quantify the expected difference between soft and hard band temperatures of the same instrument in the presence of multi temperature structure. This is shown in Fig. \ref{fig:deg3_log}, where we indicate this result from simulations by red diamonds, black squares and blue pentagons (referring to a cold component with $\SI{1}{keV}$ and a EMR of 0.01,  $\SI{2}{keV}$ and a EMR of 0.05 and  $\SI{1}{keV}$ and a EMR of 0.05, respectively), i.e. these new symbols represent the new expected ``zero-points'' given the multiple temperature components. 

We conclude that the multi temperature structure has a strong influence on the results of this test and prevents firm conclusions on the absolute calibration.
For a cold component with a temperature of $\sim\SI{1}{keV}$ and EMR higher than $\num{0.01}$, the EPIC-PN instrument seems to be in agreement with the simulations.

We want to emphasize that the non-detection of an effect of multi temperature ICM on comparing different instruments (as suggested in Section \ref{sec:disc}) is unrelated to multi temperature effects on the soft and hard band temperatures of the same instrument presented here, since the multi temperature influence may be much stronger. 
\clearpage
\section{XMM-Newton and Chandra cross calibration formulae}\label{sec:relations}
\begin{figure}[!h]
\centering
  \resizebox{0.95\hsize}{!}{\includegraphics{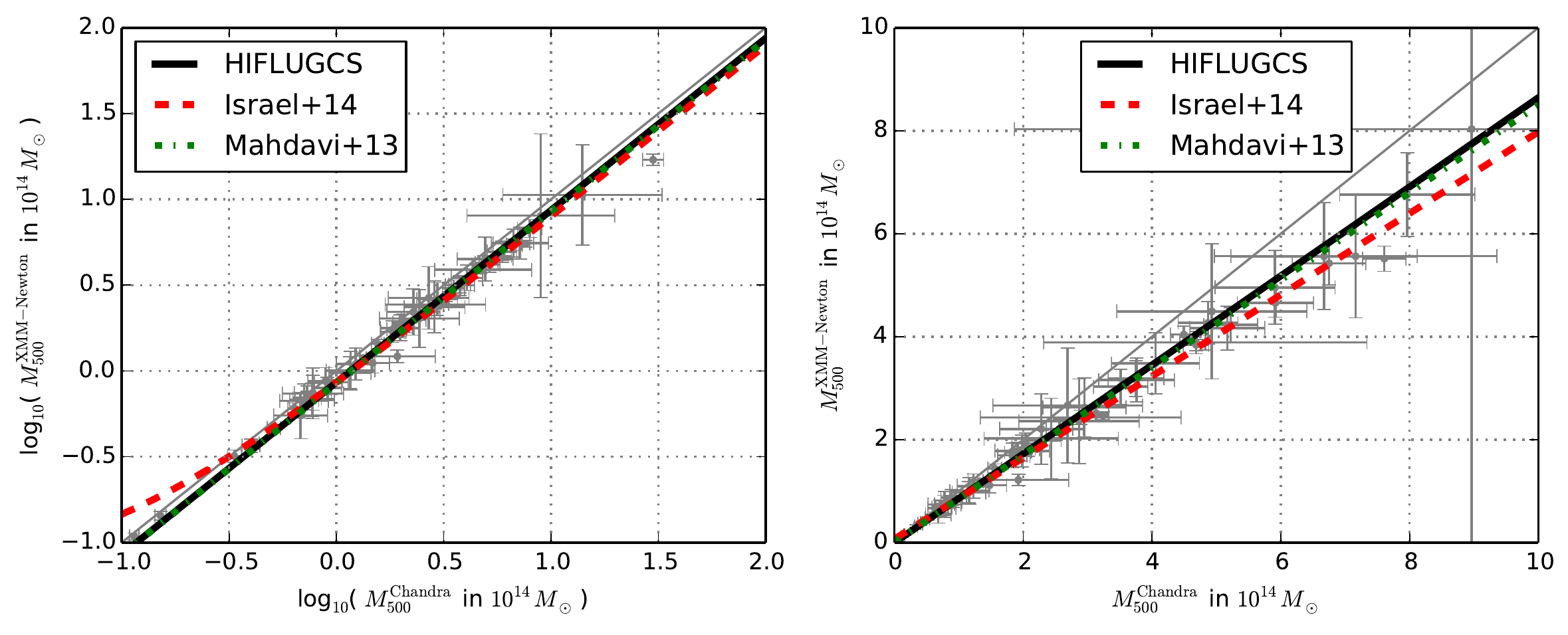}}
  \caption{Gray circles: Hydrostatic masses for Chandra (derived from data) and XMM-Newton (from rescaled Chandra temperature profiles). A best-fit relation (derived in logspace) is shown in black (Eq. \ref{eq:masses}). Also the results from \citealp{2014arXiv1408.4758I} (red dashed line) and  \citealp{2013ApJ...767..116M} (green dotted dashed line). }
  \label{fig:masses}
\end{figure}

There are several methods to perform a cross calibration between two instruments like Chandra/ACIS and XMM-Newton/EPIC using galaxy clusters. In this work we show:
\begin{itemize}
\item A correction formula for the effective area obtained using the stacked residuals ratio method (Section \ref{ch:stackedresiduals} and Table \ref{tab:spline}). For example to rescale the effective area of ACIS to give EPIC-PN consistent temperatures, one has to multiply the ACIS effective area by the spline interpolation of the ACIS/PN column of table \ref{tab:spline}.
\item A correction formula for the best-fit ICM temperature (Eq. \ref{eq:quadfunc} and Table \ref{tab:fitparam}). E.g., for an ACIS-PN conversion in the full energy band, one has to use $a=0.836$ and $b=0.016$ for the parameters in Eq. \ref{eq:quadfunc} (with ACIS as $X$ and PN as $Y$).
\item A linear relation for the hydrostatic masses (see below). The Chandra masses were calculated using the temperature and surface brightness profiles of the HIFLUGCS clusters, while the XMM-Newton masses where adopted from the rescaled Chandra temperature profiles (using Eq. \ref{eq:quadfunc} and Table \ref{tab:fitparam}).
\end{itemize}
The hydrostatic masses are shown in Figure \ref{fig:masses}. We found the following relation:
\begin{equation}
\label{eq:masses2}
M^{\rm XMM}_{500}  = 0.859^{{}+0.017}_{{}-0.016} \cdot \left( M^{\rm Chandra}_{500} \right)^{\num{1.00(2)}}  
\end{equation}
This result is in agreement with the derived relations in \cite{2013ApJ...767..116M} and also in rough agreement with \cite{2014arXiv1408.4758I}. 
In \cite{2014arXiv1408.4758I} the authors obtained Chandra masses by using Chandra cluster temperatures assuming temperature profiles (from a scaling relation from \citealp{2013SSRv..177..195R}), while the XMM-Newton masses are calculated from the rescaled temperature profiles following the temperature scaling relation presented in this work. In \cite{2013ApJ...767..116M} the authors find similar results concerning the temperature differences and also give a conversion for the hydrostatic masses between XMM-Newton and Chandra (see Fig. \ref{fig:masses}). 

The three methods presented in this work to convert between XMM-Newton and Chandra are obviously not exactly equivalent in the sense of cosmological results.

\end{appendix}
\end{document}